\documentclass[journal]{IEEEtran}
\usepackage{amsmath,amscd}
\usepackage{graphicx}
\usepackage{algorithm}
\usepackage{algorithmic}
\usepackage{amssymb,array}
\usepackage{multirow}
\usepackage{multicol}
\usepackage{enumitem}
\usepackage{bm}
\usepackage{xcolor}
\usepackage[font=small]{caption}
\usepackage{subcaption}
\usepackage{url}
\DeclareMathOperator*{\argmin}{arg\,min}
\DeclareMathOperator*{\argmax}{arg\,max}
\def\x{\bm{x}}
\def\y{\bm{y}}
\def\n{\bm{n}}

\def\A{\bm{A}}
\def\O{\bm{O}}

\def\R{\bm{R}}
\def\Q{\bm{Q}}

\def\r{\bm{r}}
\def\W{\bm{W}}
\def\I{\bm{I}}
\def\V{\bm{V}}
\def\Ee{\bm{E}}
\def\U{\bm{U}}
\def\Alpha{\bm{\alpha}}

\def\X{\bm{X}}
\def\D{\bm{D}}

\def\S{\bm{S}}
\newcommand{\E}{\mathrm{E}}
\makeatletter
\newcolumntype{"}{@{\hskip\tabcolsep\vrule width 1pt\hskip\tabcolsep}}
\makeatother

\newcommand{\mr}{\mathrm}
\newcommand{\BE}{\begin{equation}}
\newcommand{\EE}{\end{equation}}
\newcommand{\BS}{\begin{subequations}}
\newcommand{\ES}{\end{subequations}}
\ifCLASSINFOpdf
\else
\fi

\begin{document}
%
\title{Denoising-based Turbo Compressed Sensing}
%
%
%

\author{Zhipeng~Xue,
        Junjie~Ma, and~Xiaojun~Yuan,~\IEEEmembership{Senior~Member,~IEEE}
\thanks{Z. Xue and X. Yuan are with the School of Information Science and Technology, ShanghaiTech University. J. Ma is with the Department of Statistics, Columbia University. The work in this paper was partially presented at the GlobalSIP in Dec. 2016; see reference \cite{xue2016d}.}
}

\maketitle

\begin{abstract}
Turbo compressed sensing (Turbo-CS) is an efficient iterative algorithm for sparse signal recovery with partial orthogonal sensing matrices. In this paper, we extend the Turbo-CS algorithm to solve compressed sensing problems involving more general signal structure, including compressive image recovery and low-rank matrix recovery. A main difficulty for such an extension is that the original Turbo-CS algorithm requires prior knowledge of the signal distribution that is usually unavailable in practice. To overcome this difficulty, we propose to redesign the Turbo-CS algorithm by employing a generic denoiser that does not depend on the prior distribution and hence the name denoising-based Turbo-CS (D-Turbo-CS). We then derive the extrinsic information for a generic denoiser by following the Turbo-CS principle. Based on that, we optimize the parametric extrinsic denoisers to minimize the output mean-square error (MSE). Explicit expressions are derived for the extrinsic SURE-LET denoiser used in compressive image denoising and also for the singular value thresholding (SVT) denoiser used in low-rank matrix denoising. We find that the dynamics of D-Turbo-CS can be well described by a scaler recursion called MSE evolution, similar to the case for Turbo-CS. Numerical results demonstrate that D-Turbo-CS considerably outperforms the counterpart algorithms in both reconstruction quality and running time.
\end{abstract}


%
\IEEEpeerreviewmaketitle

\section{Introduction}
Compressed sensing (CS) \cite{Donoho2006} is a new paradigm for sparse signal reconstruction. A common approach for compressed sensing problem is to solve a mixed $l_1$-norm and $l_2$-norm minimization problem via convex programming \cite{tibshirani1996regression}. However, a convex program in general involves polynomial-time complexity, which causes a serious scalability problem for mass data applications.

Approximate algorithms have been extensively studied to reduce the computational complexity of sparse signal recovery. Existing approaches include match pursuit \cite{mallat1993matching}, orthogonal match pursuit \cite{tropp2004greed}, iterative soft thresholding \cite{nowak2007gradient}, compressive sampling matching pursuit \cite{needell2009cosamp}, and approximate message passing (AMP) \cite{Donoho2009}. In particular, AMP is a fast-convergence iterative algorithm based on the principle of message passing. It has been shown that, when the sensing matrix is independent and identically distributed (i.i.d.) Gaussian, AMP is asymptotically optimal as the dimension of the state space goes to infinity \cite{Bayati2011}. Also, the iterative process of AMP can be tracked through a scalar recursion called state evolution.


In many applications, compressive measurements are taken from a transformed domain, such as discrete Fourier transform (DFT), discrete cosine transform (DCT), and wavelet transform, etc. This, on one hand, can exempt us from storing the sensing matrix in implementation; on the other hand, these orthogonal transforms can be realized using fast algorithms to reduce the computational complexity. However, the AMP algorithm, when applied to orthogonal sensing, does not perform well and its simulated performance deviates away from the prediction by the state evolution.

Turbo compressed sensing (Turbo-CS) \cite{ma2014turbo} solved the above discrepancy by a careful redesign of the message passing algorithm. The Turbo-CS algorithm consists of two processing modules: One module handles the linear measurements of the sparse signal based on the linear minimum mean-square error (LMMSE) principle and calculates the so-called extrinsic information to decorrelate the input and output estimation errors; the other module combines its input with the signal sparsity by following the minimum mean-square error (MMSE) principle and also calculates the extrinsic information. The two modules are executed iteratively to refine the estimates. This is similar to the decoding process of a turbo code \cite{berrou1996near}, hence the name Turbo-CS. It has been shown that Turbo-CS considerably outperforms its counterparts for compressed sensing in both complexity and convergence speed.

In this paper, we extend the Turbo-CS algorithm to solve compressed sensing problems with partial orthogonal sensing matrices involving more general signal structures, such as compressive image recovery and low-rank matrix recovery. An immediate obstacle for such an extension is that the MMSE module in the Turbo-CS algorithm in \cite{ma2014turbo} requires the prior knowledge of the signal distribution, while the latter is generally unavailable in the new problems under concern. To overcome this obstacle, we replace the MMSE module in Turbo-CS by a generic denoiser that does not depend on the prior distribution. We derive the extrinsic information for a generic denoiser by following the Turbo-CS principle. Interestingly, we show that the resulting extrinsic denoiser falls into the category of divergence-free denoisers in \cite{ma2017oamp}. Based on that, we propose to optimize the parametric extrinsic denoisers to minimize the output mean-square error (MSE). Explicit expressions are derived for the extrinsic SURE-LET denoiser used in image denoising \cite{blu2007sure} and also for the singular value thresholding (SVT) denoiser used in low-rank matrix denoising \cite{candes2013unbiased}.

We find the dynamics of denoising-based Turbo-CS (D-Turbo-CS) can be characterized by a scaler recursion called MSE evolution. We also study the impact of the choice of the sensing matrix on the accuracy of the MSE evolution in Turbo-CS. We show that when the signals to be recovered are i.i.d., the output error of the LMMSE module can be modelled as an additive Gaussian noise and the corresponding state evolution is accurate. However, the state evolution is not necessarily accurate when correlated signals are involved, e.g., in the case of image denoising where the neighbouring pixels of an image are usually continuous in value and so are correlated to each other. We show that this problem can be solved by an appropriate design of the sensing matrix. A simple solution is to right-multiply the sensing matrix by an extra diagonal matrix with random +1 or -1 in the diagonal. This extra diagonal matrix randomly flips the signs of the signals, and effectively decorrelates the signals.

We further compare the performance of D-Turbo-CS with the state-of-the-art algorithms in the literature. For example, denoising-based AMP (D-AMP) was studied in \cite{metzler2016denoising}, and a number of popular image denoisers were examined therein. Also, the EM-GM-AMP algorithm proposed in \cite{vila2013expectation} can be applied to the compressed image denoising problem under concern. Numerical results demonstrate that D-Turbo-CS considerably outperforms D-AMP and EM-GM-AMP in both convergence rate and recovery accuracy. 

The remainder of the paper proceeds as follows. Section \ref{section_pre} takes a brief review of the Turbo-CS algorithm in \cite{ma2014turbo}. Section \ref{section3} describes how to extend the Turbo-CS algorithm for a generic denoiser. The construction of an extrinsic denoiser is discussed in Section \ref{denoisers}. Section \ref{Sec:SE-D-TurboCS} studies the MSE evolution for D-Turbo-CS. Numerical comparisons of Turbo-CS with its counterparts are presented in Section \ref{Sec:numerical}. Section \ref{conclusion} concludes the paper.
 

\section{Preliminaries}\label{section_pre}
\subsection{Compressed Sensing}
Consider the following real-valued linear system:
\BE
\y=\A\x+\n\label{linsys}
\EE
where $\x\in \mathbb{R}^n$ is an unknown signal vector, $\A\in \mathbb{R}^{m\times n}$ is a known constant matrix, and $\n$ is a white Gaussian noise vector with zero mean and covariance $\sigma^2\I$. Here, $\I$ represents the identity matrix of an appropriate size. Our goal is to recover $\x$ from the measurement $\y$. In particular, this problem is known as compressed sensing when $m< n$ and $\x$ is sparse.

Basis Pursuit De-Noising (BPDN) is a well-known approach to the recovery of $\x$ in compressed sensing, with the problem formulated as
\begin{align}
	\hat \x=\argmin_{\x\in \mathbb{R}^N}\frac{1}{2}\|\y-\A\x\|_2^2+\lambda\|\x\|_1.\label{bpdn}
\end{align}
where $\|\x\|_p=(\sum_n |x_n|^p)^{1/p}$ represents the $l_p$-norm, $x_n$ is the $n$th entry of $\x$, and $\lambda$ is a regularization parameter.
This problem can be solved by convex programming algorithms, such as the interior point method \cite{nemirovski2008interior} and the proximal method \cite{parikh2014proximal}. Interior point method has cubic computational complexity, which is too expensive for high-dimensional applications such as imaging. Proximal methods have low per-iteration complexity. However, its convergence speed is typically slow.

Message passing is a promising alternative to solve the BPDN problem in (\ref{bpdn}). To apply message passing, we first notice that (\ref{bpdn}) can be viewed as a maximum a posteriori probability (MAP) estimation problem. Specifically, we assign a prior distribution $p(\x)\propto\exp(\frac{-\lambda \|\x\|_1}{\sigma^2})$ to $\x$. Then, it is easy to verify that $\hat \x$ in (\ref{bpdn}) is equivalent to:
\begin{align}
	\begin{split}
		\hat \x=\argmax_{\x\in \mathbb{R}^n} p(\x|\y)\label{map}
	\end{split}
\end{align}
where $p(\x|\y)=p(\y|\x)p(\x)/p(\y)$.
In \cite{donoho2010message}, a factor graph was established to represent above the probability model, based on which approximate message passing (AMP) was used to iteratively solve the inference problem in (\ref{map}). As the established factor graph is dense in general, directly applying message passing to the graph leads to high complexity. To reduce complexity, two approximations are introduced in AMP: First, messages from factor nodes to variable nodes are nearly Gaussian; second, messages from variable nodes to factor nodes can be calculated by using Taylor-series approximation to reduce computational cost. It was shown in \cite{donoho2010message} that the approximation error vanishes when $m,n\rightarrow \infty$ with a fixed ratio.

The convergence of the AMP algorithm requires that the elements of the sensing matrix $\A$ are sufficiently random. It was shown in \cite{Bayati2011} that AMP is asymptotically optimal when $\A$ is i.i.d. Gaussian and the behavior of AMP can be characterized by a scaler recursion called state evolution.
\vspace{-0.5em}
\subsection{Turbo Compressed Sensing}\label{sec-turbo}
\begin{figure}[!ht]
\vspace{-1em}
	\centering
	\includegraphics[width=\columnwidth]{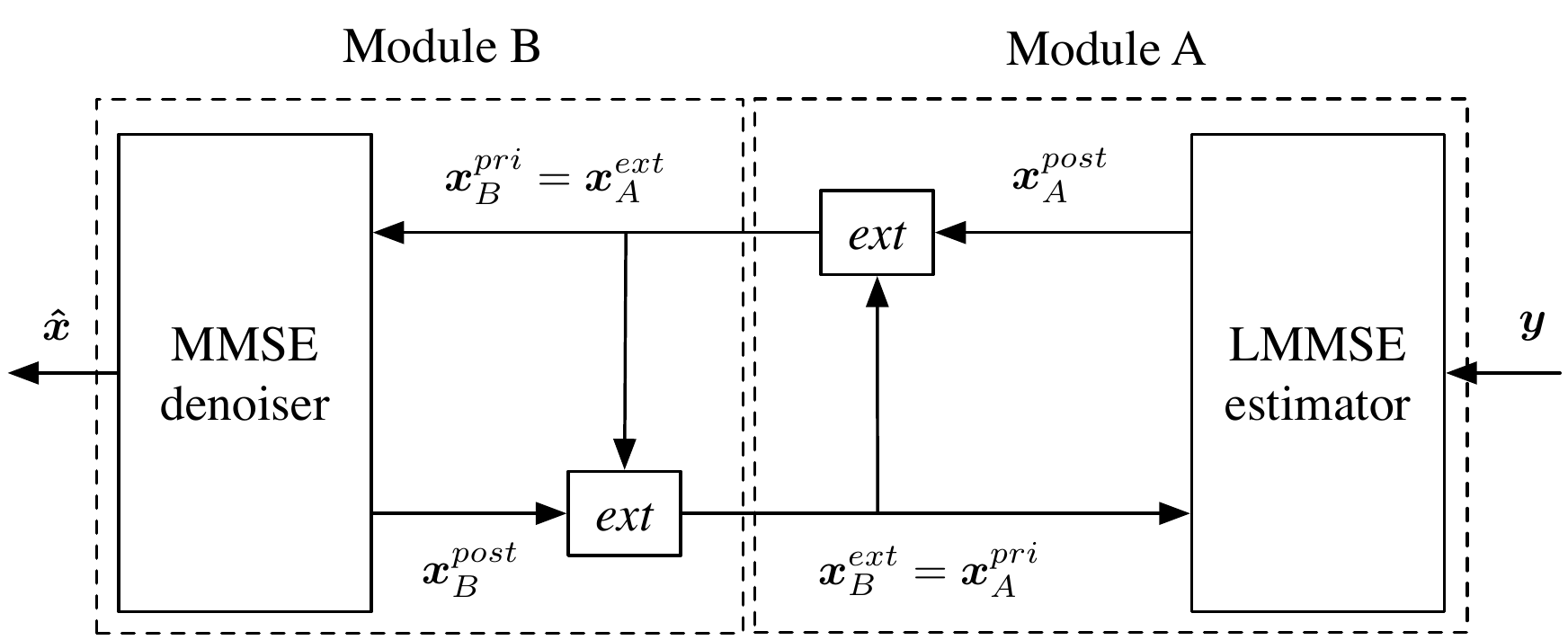}
	\caption{An illustration of the Turbo-CS algorithm proposed in \cite{ma2014turbo}.}\vspace{-0.5em}
	\label{fig1}
\end{figure}
In many applications, the sensing matrix $\A$ is neither i.i.d. nor Gaussian. 
For example, to reduce storage and computational complexity, measurements are usually taken from an orthogonal transform domain, such as DFT or DCT. In these scenarios, The performance of AMP deteriorates and the convergence of AMP is not guaranteed. This motivates the development of the Turbo-CS algorithm \cite{ma2014turbo} described below.

\begin{algorithm}
 \caption{Turbo-CS Algorithm}\label{algorithm1}
 \begin{algorithmic}[1]
 \renewcommand{\algorithmicrequire}{\textbf{Input:}}
\renewcommand{\algorithmicensure}{\textbf{Output:}}
 \renewcommand{\thealgorithm}{}
 \REQUIRE $\A, \y, \sigma^2, \x_A^{pri}=\bm{0}$
 \\
\WHILE{the stopping criterion is not met}
\STATE $\x_A^{ext}=\x_A^{pri}+\frac{n}{m}\A^{T}(\y-\A\x_A^{pri})$ \%Module A
\STATE $v_A^{ext}=\left(\frac{n}{m}-1\right)v_A^{pri}+\frac{n}{m}\sigma^2$
\STATE $\x_B^{pri}=\x_A^{ext},v_B^{pri}=v_A^{ext}$
\STATE $x_{B,i}^{post}=\E\left[x_i|x_{B,i}^{pri}\right]$ \%Module B
\STATE $v_B^{post}=\frac{1}{n}\sum_{i=1}^n \text{var}\left[x_i|x_{B,i}^{pri}\right]$
\STATE $v_A^{pri}=v_B^{ext}=\left(\frac{1}{v_B^{post}}-\frac{1}{v_B^{pri}}\right)^{-1}$
\STATE $\x_A^{pri}=\x_B^{ext}=v_B^{ext}\left(\frac{\x_B^{post}}{v_B^{post}}-\frac{\x_B^{pri}}{v_B^{pri}}\right)$
 \ENDWHILE
  \ENSURE  $\x_B^{post}$
 \end{algorithmic}
 \end{algorithm}
The block diagram of the Turbo-CS algorithm is illustrated in Fig. \ref{fig1}. Turbo-CS bears a structure similar to a turbo decoder \cite{berrou1996near}, hence the name Turbo-CS. As illustrated in Fig. \ref{fig1}, the Turbo-CS algorithm consists of two modules. Module A is basically a linear minimum mean square error (LMMSE) estimator of $\x$ based on the measurement $\y$ and the messages from Module B. Module B performs minimum mean square error (MMSE) estimation that combines the prior distribution of $\x$ and the messages from Module A. The two modules are executed iteratively to refine the estimate of $\x$. The detailed operations of Turbo-CS are presented in Algorithm \ref{algorithm1}.

We now give more details of Algorithm \ref{algorithm1}. Module A estimates $\x$ based on the measurement $\y$ in (\ref{linsys}) with $\x$ \textit{a priori} distributed as $\x \sim \mathcal{N}(\x_A^{pri},v_A^{pri}\I)$. Given $\y$ with $\x \sim \mathcal{N}(\x_A^{pri},v_A^{pri}\I)$, the posterior distribution of each $x_i$ is still Gaussian with posterior mean and variance given by \cite{kay2013fundamentals}
\BS
\begin{align}
	x_{A,i}^{post}&=x_{A,i}^{pri}+\frac{v_A^{pri}}{v_A^{pri}+\sigma^2}\bm{a}_{i}^T(\y-\A\x_A^{pri})\\
v_A^{post}&=v_A^{pri}-\frac{m}{n}\frac{(v_A^{pri})^2}{v_A^{pri}+\sigma^2},
\end{align}\label{apost}\ES
where $\bm{a}_i$ is the $i$th column of $\A$. Note that as the measurement $\y$ is linear in $\x$, the \textit{a posteriori} mean in (\ref{apost}a) is also called the LMMSE estimator of $x_i$.

The posterior distributions cannot be used directly in message passing due to the correlation issue. Instead, we need to calculate the so-called extrinsic message \cite{berrou1996near} for each $x_i$ by excluding the contribution of the input message of $x_i$. That is, the extrinsic distribution of each $x_i$ satisfies
\begin{align}
	\begin{split}
		\mathcal{N}_{x_i}(x_{A,i}^{pri},v_{A,i}^{pri})\mathcal{N}_{x_i}(x_{A,i}^{ext},v_{A,i}^{ext})\doteq\mathcal{N}_{x_i}(x_{A,i}^{post},v_{A,i}^{post}),\label{extcondition}
	\end{split}
\end{align}
where $\mathcal{N}_{x}(m,v)=\frac{1}{\sqrt{2\pi v}}\exp(-\frac{1}{2v}(x-m)^2)$, and ``$\doteq$" represents equality up to a constant multiplicative factor. From (\ref{extcondition}), the extrinsic mean and variance of $x_i$ are respectively given in \cite{guo2011concise} as
\BS
\begin{align}
		x_{A,i}^{ext}&=v_A^{ext}\left(\frac{x_{A,i}^{post}}{v_A^{post}}-\frac{x_{A,i}^{pri}}{v_A^{pri}}\right)\\
v_A^{ext}&=\left(\frac{1}{v_A^{post}}-\frac{1}{v_A^{pri}}\right)^{-1}.
\end{align}\label{aext}\ES
Combining (\ref{apost}) and (\ref{aext}), we obtain Lines 2 and 3 of Algorithm \ref{algorithm1}.

It is worth noting that (\ref{extcondition}) implies the independence of the input distortion $x_{A,i}^{pri}-x_i$ and the output distortion $x_{A,i}^{ext}-x_i$. Further more, for Gaussian distributions, independence is equivalent to uncorrelatedness. Thus, we have
 \begin{align}
 	\begin{split}
 		\E\left[(x_i-x_{A,i}^{pri})(x_i-x_{A,i}^{ext})\right]=0,
 	\end{split}\label{uncorrelation}
 \end{align}
where the expectation is taken over the joint probability distribution of $x_i$, $x_{A,i}^{pri}$, and $x_{A,i}^{ext}$.

We now consider Module B. Recall that Module B estimates each $x_i$ by combining  the prior distribution $x_i\sim p(x_i)$ and the message from Module A. Note that the message $x_{A,i}^{ext}$ from Module A is now treated as an input of Module B, denoted by $x_{B,i}^{pri}$. Following \cite{ma2014turbo}, we model each $x_{B,i}^{pri}$ as an observation of $x_i$ corrupted by an additive noise:
\begin{align}
	x_{B,i}^{pri}=x_i+n_{B,i}^{pri}\label{xbpri}
\end{align}
where $n_{B,i}^{pri}\sim\mathcal{N}(0,v_B^{pri})$ is independent of $x_i$. The \textit{a posteriori} mean and variance of each $x_i$ for Module B are respectively given by
\begin{align}
	\begin{split}
	x_{B,i}^{post}&=\E[x_i|x_{B,i}^{pri}]\\
	v_B^{post}&=\frac{1}{n}\sum_{i=1}^n \text{var}[x_i|x_{B,i}^{pri}],
	\end{split}
\end{align}
where $\text{var}[x|y]$ denotes conditional variance of $x$ given $y$. Similar to (\ref{aext}), the extrinsic variance and mean of $\x$ for Module B are respectively given by Lines 7 and 8 of Algorithm \ref{algorithm1}. Also, similar to (\ref{uncorrelation}), the extrinsic distortion is uncorrelated with the prior distortion, i.e. 
 \begin{align}
 	\begin{split}
 		\E\left[(x_i-x_{B,i}^{pri})(x_i-x_{B,i}^{ext})\right]=0,
 	\end{split}\label{uncorrelationB}
 \end{align}
 where the expectation is taken over the joint propability distribution of $x_i$, $x_{B,i}^{pri}$ and $x_{B,i}^{ext}$. Later, we will see that (\ref{uncorrelationB}) plays an important role in the extension of Turbo-CS.

\section{Denoising-based Turbo CS}\label{section3}
\subsection{Problem Statement}
In Algorithm \ref{algorithm1}, the operation of Module B requires the knowledge of the prior distribution of $\x$. However, such prior information is difficult to acquire in many applications. Low-complexity robust denoisers, rather than the optimal MMSE denoiser, are usually employed in practice, even when the prior distribution of $\x$ is available.

\begin{figure}[!ht]
\vspace{-1em}
	\centering
	\includegraphics[width=\columnwidth]{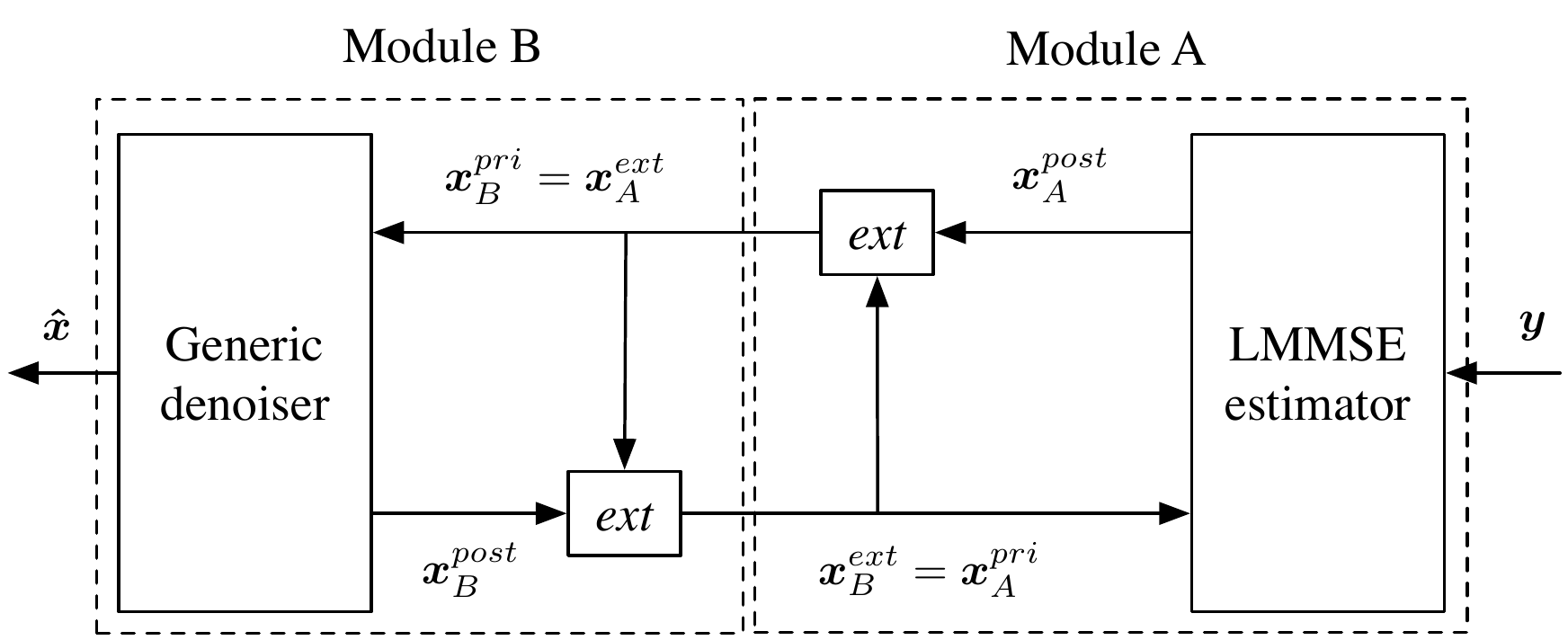}
	\caption{An illustration of the denoising-based Turbo-CS algorithm.}
	\label{fig2}
	\vspace{-1.5em}
\end{figure}

Turbo-CS with a generic denoiser is illustrated in Fig. \ref{fig2}. Compared with Fig. \ref{fig1}, the only difference is that Turbo-CS in Fig. \ref{fig2} replaces the MMSE denoiser by a generic denoiser, defined as
\begin{align}
	\begin{split}
		\x_B^{post}=\D(\x_B^{pri};v_B^{pri}, \bm{\theta}),
	\end{split}
\end{align} 
where $\D(\cdot)$ represents the denoising function with $\x_B^{pri}$ being the input, $\x_B^{post}$ being the output, and $v_B^{pri}$ and $\bm{\theta}$ being the parameters. Note that the choice of $\bm{\theta}$ will be specified when a specific denoiser is involved. For brevity, we may simplify the notation $\D(\x_B^{pri};v_B^{pri},\bm{\theta})$ to $\D(\x_B^{pri})$ in circumstances without causing ambiguity. Also, we denote the $i$th entry of $\x_B^{post}$ as
\begin{align}
	\begin{split}
		x_{B,i}^{post}=D_i(\x_B^{pri}).
	\end{split}\label{xB}
\end{align}

With the above replacement, the main challenge is how to calculate the extrinsic message of each $x_i$ for Module B, without the prior knowledge of the distribution $p(x_i)$. Note that Lines 7 and 8  of Algorithm \ref{algorithm1} cannot be used any more since they hold only for the MMSE denoiser.

\subsection{Extrinsic Messages for a Generic Denoiser}
We now describe how to calculate the extrinsic messages for a generic denoiser. Without loss of generality, denote the extrinsic output of Module B by $\x_B^{ext}=\D^{ext}(\x_B^{pri})$. We call $\D^{ext}(\x_B^{pri})$ a extrinsic denoiser. Similarly to Line 8 of Algorithm \ref{algorithm1}, we construct $\x_B^{ext}$ by a linear combination of the \textit{a priori} mean and the \textit{a posteriori} mean: 
\BE
\x_B^{ext}=\D^{ext}(\x_B^{pri})=c(\x_B^{post}-\alpha \x_B^{pri})\label{eqext}
\EE
where $c$ and $\alpha$ are coefficients to be determined. Clearly, (\ref{eqext}) is identical to Line 8 of Algorithm \ref{algorithm1} by letting $c=\frac{v_B^{ext}}{v_B^{post}}$ and $\alpha=\frac{v_B^{post}}{v_B^{pri}}$. Here, we require that $c$ and $\alpha$ are chosen such that

\begin{enumerate}[label=(\roman*),align=left]
	\item The extrinsic distortion is uncorrelated with the prior distortion, i.e.\label{cond1}
	\begin{align}
		\E[(\x-\x_B^{pri})^T(\x-\x_B^{ext})]=0;\label{relexcond}
	\end{align}
	\item $\E[\|\x_B^{ext}-\x\|^2]$ is minimized. \label{cond2}
\end{enumerate}

From the discussions in Section \ref{sec-turbo}, the calculation of the extrinsic messages in Lines 8 and 9 satisfies the above two conditions when the MMSE denoiser is employed. Note that (\ref{relexcond}) is a relaxation of (\ref{uncorrelationB}) since (\ref{uncorrelationB}) implies (\ref{relexcond}) but the converse does not necessarily hold. Later we will see that this relaxation is good for many applications. What remains is to determine $c$ and $\alpha$ satisfying conditions \ref{cond1} and \ref{cond2} for a generic denoiser. This is elaborated in the following.
\subsubsection{Determining parameter $\alpha$}
As mentioned in Section \ref{sec-turbo}, the input message of Module B can be modeled by (\ref{xbpri}), where the noise part $n_{B,i}^{pri}$ is independent of $x_i$. Then
\BS
\begin{align}
	\E[(\x-\x_B^{pri})^T(\x-\x_B^{ext})]&= \E[(\n_B^{pri})^T(\x-\x_B^{ext})]\\&=\sum_{i=1}^n \E\left[n_{B,i}^{pri} x_{B,i}^{ext}\right]\label{equnco}
\end{align}\label{relaxu}\ES
where (\ref{relaxu}a) follows from (\ref{xbpri}), and (\ref{relaxu}b) follows by noting $\E[(\n_B^{pri})^T \x]=0$. 
To proceed, we introduce the Stein's lemma \cite{stein1981estimation} as follows: For a normally distributed random variable $y\sim \mathcal{N}(\mu_{y},\sigma_{y}^2)$, and a differentiable function $h: \mathbb{R}\rightarrow \mathbb{R}$ such that $\E[|h'(y)|]<\infty$, we have
\BE
\sigma_{y}^2\E[h'(y)]=\E[(y-\mu_{y})h(y)].
\EE
Then
\BS
\begin{align}
	\E\left[n_{B,i}^{pri}x_{B,i}^{ext}\right]=&c \E\left[n_{B,i}^{pri}(x_{B,i}^{post}-\alpha(x_i+n_{B,i}^{pri}))\right]\\\notag
	=&c\E[n_{B,i}^{pri} x_{B,i}^{post}]-c\alpha \E[n_{B,i}^{pri}  x_i]
	\\& -c\alpha \E[n_{B,i}^{pri}n_{B,i}^{pri} ]\\
	=&c\E \left[n_{B,i}^{pri} D_i(\x_B^{pri})\right ]-c\alpha  v_B^{pri}\\
		=&c\E\left[n_{B,i}^{pri} D_i(\x+\n_B^{pri})\right]- c\alpha  v_B^{pri}\\
		=&cv_B^{pri} \E\left[D_i'(\x+\n_B^{pri})\right]-c\alpha  v_B^{pri}
\end{align}\label{enx}\ES
where $D_i'(\x+\n_B^{pri})$ denotes the partial derivative of $D_i(\x+\n_B^{pri})$ with respect to variable $n_{B,i}^{pri}$ and the expectation is taken over the joint probability distribution of $\n_{B}^{pri}$ and $\x$. In the above, (\ref{enx}a) follows from (\ref{xbpri}) and (\ref{eqext}), (\ref{enx}c) from $\E[x_{B,i}^{pri}x_i]=0$ and $\E[n_{B,i}^{pri} n_{B,i}^{pri}]=v_B^{pri}$, and (\ref{enx}e) from the Steins's lemma by letting $y=n_{B,i}^{pri}$. Combining (\ref{relexcond}), (\ref{equnco}), and (\ref{enx}), we obtain
\BS
\begin{align}
	\alpha &= \frac{1}{n}\E\left[\sum_{i=1}^n D_i'(\x_B^{pri})\right]\\&\approx\frac{1}{n}\sum_{i=1}^n D_i'(\x_B^{pri})\\
	&=\frac{1}{n}\mr{div}\{\D(\x_B^{pri})\},
\end{align}\label{alpha}\ES
where $\mr{div}$ denotes divergence, and $D_i'(\x_B^{pri})$ is the partial derivative of $D_i(\x_B^{pri})$ with respect to $x_{B,i}^{pri}$. Note that the approximation in (\ref{alpha}b) becomes accurate when $n$ is large. Also, with this approximation, the calculation of $\alpha$ does not depend on the distribution of $\x$.

By substituting (\ref{alpha}) into (\ref{eqext}), we obtain
\BS
\begin{align}
		\x_B^{ext}&=c\left(\x_B^{post}-\frac{1}{n}\mr{div}\{\D(\x_B^{pri})\} \x_B^{pri}\right)\\
		&=c\left(\D(\x_B^{pri})-\frac{1}{n}\mr{div}\{\D(\x_B^{pri})\} \x_B^{pri}\right)\\
		&=\D^{ext}(\x_B^{pri}),
\end{align}\label{eqext2}\ES
where the extrinsic denoiser $\D^{ext}(\cdot)$ is defined as
\begin{align}
	\begin{split}
		\D^{ext}(\r)=c\left(\D(\r)-\frac{1}{n}\mr{div}\{\D(\x_B^{pri})\} \r\right).
	\end{split}\label{eqext3}
\end{align}
The divergence of $\D^{ext}(\r)$ at $\r=\x_B^{ext}$ is zero by noting
\begin{align} \label{Eqn:divergence-free-2b}
\begin{split}
\mr{div}\{\D^{ext}(\x_B^{pri})\}&\!=\!c\!\left( \mr{div}\{\D(\x_B^{pri})\}  \!-\!\mr{div}\{\D(\x_B^{pri})\}  \right)\\&\!=\!0.
\end{split}
\end{align}
Thus, $\D^{ext}(\cdot)$ belongs to the family of divergence-free denoisers proposed in \cite{ma2017oamp}.
\subsubsection{Determining parameter $c$} Ideally, we want to choose parameter $c$ to satisfy condition \ref{cond2} below (\ref{relexcond}). However, the MSE is difficult to evaluate as the distribution of $\x$ is unknown. To address this problem, we use the Stein's unbiased risk estimate (SURE) \cite{stein1981estimation} to approximate the MSE.

To be specific, consider the signal model
\BE
\bm{r}=\bm{x}+\tau\bm{n},
\EE
where $\bm{n}\in \mathbb{R}^{n\times 1}$ is the additive Gaussian noise draw from $\mathcal{N}(0,\I)$. The mean square error of denoiser $\D(\r)$ is defined by
\BE \label{Eqn:MSE_def}
\mr{MSE}=\frac{1}{n}\E\left[\|\bm{D}(\bm{r})-\bm{x}\|^2\right].
\EE
The SURE of the MSE of $\D(\r)$ is given by
\BE \label{Eqn:SURE_def}
\widehat{\mr{MSE}}=\frac{1}{n}\|\D(\r)-\r\|^2+\frac{2\tau^2}{n}\text{div}\{\D(\r)\}-\tau^2.
\EE
Compared with the MSE in \eqref{Eqn:MSE_def}, the SURE in (\ref{Eqn:SURE_def}) does not involve the distribution of $\x$. We next use SURE as a surrogate for MSE and tune the denoiser by minimizing the SURE. Recall from (\ref{xbpri}) that $\x_B^{pri}$ can be represented as $\x_B^{pri}=\x+\n_B^{pri}$. Let $\tau=\sqrt{v_B^{pri}}$. Then, applying (\ref{Eqn:SURE_def}) to $\D^{ext}(\x_B^{pri})$, we obtain
\begin{align}
\begin{split}
	\widehat{\mr{MSE}}=&\frac{1}{n}\|\D^{ext}(\x_B^{pri})-\x_B^{pri}\|^2+\frac{2v_B^{pri}}{n}\text{div}\{\D(\x_B^{pri})\}-v_B^{pri}\\
	=&\frac{1}{n}\|\D^{ext}(\x_B^{pri})-\x_B^{pri}\|^2-v_B^{pri}\\
=&\frac{1}{n}\left\|c\left( \!\D(\x_B^{pri}){-}\frac{1}{n}\mr{div}\{\D(\x_B^{pri})\} \x_B^{pri} \right)\!\!{-}\x_B^{pri}\right\|^2\!{-}v_B^{pri}\label{Eqn:SURE}
\end{split}
\end{align}
where the second step follows from (\ref{Eqn:divergence-free-2b}), and the last step from (\ref{eqext2}). Minimizing the SURE given in \eqref{Eqn:SURE}, we obtain the optimal $c$ given by
\BE \label{Eqn:C_opt}
c^{opt}=\frac{(\x_B^{pri})^T\left(   \D(\x_B^{pri})-\frac{1}{n}\mr{div}\{\D(\x_B^{pri})\} \x_B^{pri}\right)}{\| \D(\x_B^{pri})-\frac{1}{n}\mr{div}\{\D(\x_B^{pri})\} \x_B^{pri}\|^2}.
\EE
 
\subsection{Denoising-based Turbo CS}
We are now ready to extend Turbo-CS for a generic denoiser. We refer to the extended algorithm as Denoising-based Turbo-CS (D-Turbo-CS). The details of D-Turbo-CS are presented in Algorithm \ref{algorithm2}.

\begin{algorithm}
 \caption{D-Turbo-CS Algorithm}
 \begin{algorithmic}[1]
 \renewcommand{\algorithmicrequire}{\textbf{Input:}}
 \renewcommand{\algorithmicensure}{\textbf{Output:}}
 \REQUIRE $\A, \y, \sigma^2, \x_A^{pri}=0$
 \\
\WHILE{the stopping criterion is not met}
  	\STATE $\x_A^{ext}=\x_A^{pri}+\frac{n}{m}\A^{T}(\y-\A\x_A^{pri})$ \%Module A
\STATE $v_A^{ext}=\left(\frac{n}{m}-1\right)v_A^{pri}+\frac{n}{m}\sigma^2$
\STATE $\x_B^{pri}=\x_A^{ext},v_B^{pri}=v_A^{ext}$ \\
\STATE $\x_B^{post}=\D(\x_B^{pri};v_B^{pri},\bm{\theta})$ \%Module B
\STATE $\x_B^{ext}=c^{opt}(\x_B^{post}+\alpha \x_B^{pri})$ 
\STATE $v_B^{ext}=\frac{\|\y-\A\x_B^{ext}\|^2-m\sigma^2}{m}$
\STATE $\x_A^{pri}=\x_B^{ext},v_A^{pri}=v_B^{ext}$
\ENDWHILE
  \ENSURE  $\x_B^{post}$
 \end{algorithmic}\label{algorithm2}
 \end{algorithm}
 
Compared with Turbo-CS, D-Turbo-CS has the same operations in Module A. But for Module B, D-Turbo-CS employs a generic denoiser, rather than the MMSE denoiser. Correspondingly, the extrinsic mean is calculated using Line 6 of Algorithm \ref{algorithm2}; the extrinsic variance is calculated in Line 7 by following Eqn. (71) in \cite{vila2013expectation}.

\section{Construction of Extrinsic Denoisers}\label{denoisers}
Various denoisers have been proposed in the literature for noise suppression. For example, the SURE-LET \cite{blu2007sure}, the BM3D \cite{dabov2006image}, and the dictionary learning \cite{elad2006image} are developed for image denoising; the singular value thresholding (SVT)  \cite{cai2010singular} is used for low-rank matrix denoising. In this section, we study the applications  of these denoisers in D-Turbo-CS. We describe how to construct the corresponding extrinsic denoiser $\D^{ext}(\r;\bm{\theta})$ for any given denoiser $\D(\r;\bm{\theta})$. Based on that, we further consider optimizing the denoiser parameter $\bm{\theta}$.

\subsection{Extrinsic SURE-LET Denoiser}
We start with the SURE-LET denoiser. A SURE-LET denoiser is constructed as a linear combination of some kernel functions. The combination coefficients are determined by minimize the SURE of the MSE \cite{blu2007sure}.

Specifically, a SURE-LET denoiser is constructed as
\BS \label{Eqn:LET}
\begin{align}
\D(\r;\bm{\theta})&=\sum_{k=1}^{K}\theta_k \O{\bm{\psi}}_k(\O^T\r)\\
&=\sum_{i=1}^K \theta_k \bm{\Psi}_k(\r),
\end{align}
\ES
where $\O\in \mathbb{R}^{n\times n}$ is an orthonormal transform matrix, $\bm{\psi}_k: \mathbb{R}^n\rightarrow \mathbb{R}^n$ for $k= 1,\cdots,K$ are kernel functions, $\bm{\theta}=[\theta_1,\theta_2,\cdots,\theta_k]^T$, and $\bm{\Psi}_k(\r)=\O\bm{\psi}_k(\O^T\bm{r})$ . $\O$ can be the Haar wavelet transform matrix or the DCT transfrom matrix.

The choice of kernel functions $\{\bm{\psi}_k\}$ depends on the structure of the input signals. For example, the authors in \cite{guo2015near} proposed the following piecewise linear kernel functions for sparse signals:
\begin{align}
\begin{split}
	\psi_{1,i}(\r)= \begin{cases}
 0  & r_i\leq-2\beta_{1},r_i\geq2\beta_{1} \\
 -\frac{r_i}{\beta_{1}}-2 & -2\beta_{1}<r_i<-\beta_{1} \\
 \frac{r_i}{\beta_{1}} &-\beta_{1}\leq r_i\leq\beta_{1}\\
 -\frac{r_i}{\beta_{1}}+2 & \beta_{1}<r_i<2\beta_{1}
\end{cases}\label{kernel1}
\end{split}
\\
\begin{split}
	\psi_{2,i}(\r)= \begin{cases}
-1 & r_i\leq-\beta_{2}\\
\frac{r_i+\beta_{1}}{\beta_{2}-\beta_{1}} & -\beta_{2}<r_i<-\beta_{1}\\
0 & -\beta_{1}\leq r_i\leq\beta_{1}\\
\frac{r_i-\beta_{1}}{\beta_{2}-\beta_{1}} & \beta_{1}<r_i<\beta_{2}\\
1 & r_i\geq\beta_{2}
\end{cases}\label{kernel2}
\end{split}\\
	\begin{split}
	\psi_{3,i}(\r)= \begin{cases}
r_i+\beta_{2} & r_i\leq-\beta_{2}\\
 0 & -\beta_{2}<r_i<\beta_{2}\\
r_i-\beta_{2} & r_i\geq\beta_{2}
\end{cases}\label{kernel3}
\end{split}
\end{align}
where $\psi_{k,i}(\r)$ represents the $i$th element of $\bm{\psi_{k}}(\r)$, $r_i$ is the $i$th element of $\r$, $\beta_1$ and $\beta_2$ are constants chosen based on the noise level $\tau^2$. The recommended values of $\beta_1$ and $\beta_2$ can be found in \cite{guo2015near}.

For SURE-LET denoiser $\D(\r;\bm{\theta})$ in (\ref{Eqn:LET}), the corresponding extrinsic denoiser $\D^{ext}(\r;\bm{\theta})$ is given by
\BS
\begin{align}
	\bm{D}^{ext}(\r;\bm{\theta}){=}&c\!\left(\sum_{i=1}^K \!\theta_k \bm{\Psi}_k(\r){-}\!\frac{1}{n}\mr{div}\!\left\{\!\sum_{i=1}^K \!\theta_k \bm{\Psi}_k(\r)\!\right\}\! \r\!\right)\\
{=}&\sum_{i=1}^K \theta_k' \left(\bm{\Psi}_k(\r)-\frac{1}{n} \mr{div}\{\bm{\Psi}_k(\r)\} \r\right),
\end{align}\label{dext}\ES
where (\ref{dext}a) is from (\ref{eqext3}), and $\theta_k'=c\theta_k$, for $k=1,\cdots,K$.

We next determine the optimal $\bm{\theta}'=[\theta_1',\cdots,\theta_K']^T$ by minimizing the SURE. From (\ref{Eqn:SURE_def}), the SURE of $\D^{ext}(\r,\bm{\theta})$ is given by 
\BS
\begin{align}
		\widehat{\mr{MSE}}\!=&\frac{1}{n}\|\D^{ext}(\r,\bm{\theta})\!-\!\r\|^2\!+\!\frac{2\tau^2}{n}\text{div}\{\D^{ext}(\r)\}-\tau^2\\
		\!=&\frac{1}{n}\!\left\|\sum_{i=1}^K \!\theta_k'\! \left(\!\bm{\Psi}_k(\r)\!{-}\frac{1}{n} \mr{div}\{\bm{\Psi}_k(\r)\} \r\!\right)\!\!-\!\r\right\|^2\!\!-\tau^2\\
		\!= &\frac{1}{n}\!\left\|\sum_{k=1}^K \!\theta_k'\!\left({\bm{\psi}}_k(\tilde{\r})\!{-}\frac{1}{n}\text{div}\{{\bm{\psi}}_k(\tilde{\r}) \}\tilde{\r}\right)\!{-}\tilde{\r}\right\|^2\!{-}\tau^2
\end{align}\label{Eqn:SURE2}\ES
where (\ref{Eqn:SURE2}b) follows from (\ref{Eqn:divergence-free-2b}) and (\ref{dext}), (\ref{Eqn:SURE2}c) follows from $\bm{\Psi}_k(\O\tilde{\r})=\O \bm{\psi}_k(\tilde{\r})$ and $\mr{div}\{\bm{\Psi}_k(\r)\}=\mr{div}\{\bm{\psi}_k(\tilde{\r})\}$ with $\tilde{\r}=\O^T\bm{r}$.

The optimal $\bm{\theta}'$ that minimizes $\widehat{\mr{MSE}}$ in (\ref{Eqn:SURE2}) is given by
\BS
\begin{align} \label{Eqn:alpha_opt_AMP}
	\begin{split}
		(\bm{\theta}')^{opt}=\bm{M}^{-1}\bm{b},
	\end{split}
\end{align}\ES
where the $(i,j)$th entry of $\bm{M}\in\mathbb{R}^{K\times K}$ and the $i$th entry of $\bm{b}\in\mathbb{R}^{K\times1}$ are respectively given by
\BS
\begin{align} 
M_{i,j}&=[\bm{\psi}_i^{ext}(\tilde{\r})]^T\bm{\psi}_j^{ext}(\tilde{\r}) \label{Eqn:opt_M_AMP}\\
b_i&=[\bm{\psi}_i^{ext}(\tilde{\r})]^T\tilde{\r},\label{Eqn:opt_b_AMP}
\end{align}\ES
with
\BE
\bm{\psi}_k^{ext}(\tilde{\r})=\bm{\psi}_k(\tilde{\r})-\frac{1}{n}\mr{div}\{\bm{\psi}_k(\tilde{\r})\}\tilde{\r}.
\EE


\subsection{Extrinsic SVT Denoiser}
In many applications, data are arranged in a matrix form. Thus, we rearrange the signal vector $\x \in \mathbb{R}^{n\times 1}$ into a matrix $\X \in \mathbb{R}^{n_1\times n_2}$ with $n_1 n_2=n$, and consider the recovery of a low-rank $\bm{X}$ from the noisy observation
\BE
\bm{R}=\bm{X}+\tau\bm{N},
\EE
where $\tau$ is the noise level, and $\bm{N}$ contains i.i.d. Gaussian noise with zero mean and unit variance. Let $r$ be the rank of $\X$. We assume that $\X$ is a low-rank matrix, i.e. $r\ll n_1,n_2$. A popular method for low-rank matrix denoising is the so-called singular value thresholding (SVT) \cite{candes2013unbiased}:
\begin{align} \label{Eqn:SVT_1}
	\begin{split}
		\text{SVT}(\R;\theta)=\argmin_{\X} \frac{1}{2}\|\R-\X\|_F^2+\theta  \|\X\|_{\ast},
	\end{split}
\end{align}
where $\theta>0$ is a regularization parameter, $\|\cdot\|_{F}$ denotes the Frobenius norm, and $\|\cdot\|_{\ast}$ denotes the nuclear norm.
The singular value decomposition of $\R$ is given by
\BE
\R=\U\bm{\Sigma} \V^T=\sum_{i=1}^r \sigma_i \bm{u}_i \bm{v}_i^T,
\EE
where $\bm{\Sigma}=\text{diag}\{\sigma_1,\sigma_2,\cdots,\sigma_r\}\in \mathbb{R}^{r\times r}$, $\U=[\bm{u}_1,\bm{u}_2,\cdots,\bm{u}_r]\in \mathbb{R}^{n_1\times r}$ satisfies $\U^T \U=\I$, and $\V=[\bm{v}_1,\bm{v}_2,\cdots,\bm{v}_r]\in \mathbb{R}^{n_2\times r}$ satisfies $\V^T \V=\I$. Then, the SVT denoiser in \eqref{Eqn:SVT_1} has the following closed-form expression \cite{candes2013unbiased}:
\begin{align}
	\begin{split}
			\text{SVT}(\R;\theta)=\sum_{i=1}^r (\sigma_i-\theta)_{+}\bm{u}_i \bm{v}_i^T.
	\end{split}\label{svt}
\end{align}
From \cite{candes2013unbiased}, the divergence of the SVT denoiser $\text{SVT}(\R;\theta)$ has a closed-form expression given by
\begin{align}
	\begin{split}
		\mr{div}\{\!\text{SVT}(\R;\theta)\!\}\!=&|n_1\!\!-\!n_2|\!\sum_{i=1}^{n_{m}}\!\left(\!1\!-\!\frac{\theta}{\sigma_i}\!\right)_{+}\!\!+\!\sum_{i=1}^{n_{m}}\!\mathbb{I}(\sigma_i\!>\!\theta)\\&+2\sum_{i\neq j,i,j=1}^{n_{m}}\frac{\sigma_i(\sigma_i-\theta)_{+}}{\sigma_i^2-\sigma_j^2}.
	\end{split}
\end{align}
where $n_{m}=\min(n_1,n_2)$. For an SVT denoiser, we construct the extrinsic denoiser $\D^{ext}(\R;\theta)$ based on (\ref{eqext3}) as
\BS
\begin{align}
&\D^{ext}(\R;\theta)\notag\\
&=c\left(\text{SVT}(\R;\theta)-\frac{1}{n}\text{div}\{\text{SVT}(\R;\theta)\}\R\right)\\
&=c\!\left(\!\U\text{SVT}(\bm{\Sigma};\theta)\V^T{-}\frac{1}{n}\mr{div}\!\{\text{SVT}(\R;\theta)\!\}\U\bm{\Sigma}\V^T\!\right)\\
&= c\U\left(\text{SVT}(\bm{\Sigma};\theta)-\frac{1}{n}\text{div}\{\text{SVT}(\R;\theta)\}\bm{\Sigma}\right)\V^T\\
&= c\U\text{diag}\left(\bm{\Phi}(\bm{\bm{\sigma}};\theta)\right) \V^T,	
\end{align}\label{svtext}\ES
where
\BS
\begin{align}
		\bm{\Phi}(\bm{\sigma};\theta)&= \!(\bm{\sigma}\!-\!\theta)_{+}\!-\!\frac{1}{n}\text{div}\{\text{SVT}(\R;\theta)\}\bm{\sigma}\in \mathbb{R}^{r\times 1}\\
		\bm{\sigma}&=[\sigma_1,\sigma_2,\cdots,\sigma_r]^T.
\end{align}
\ES

We define the MSE and the SURE of $\D^{ext}(\R;\theta)$ respectively as
\BS
\begin{align}
\mr{MSE}&=\frac{1}{n}\|\bm{D}^{ext}(\bm{R};\theta)-\bm{X}\|_F^2\\
\widehat{\mr{MSE}}&=\frac{1}{n}\|\D^{ext}\!(\R;\theta)\!\!-\!\!\R\|_F^2{+}\frac{2\tau^2}{n}\text{div}\{\!\D^{ext}(\R;\theta)\!\}{-}\tau^2\label{Eqn:SURE_DF_M}
\end{align}
where the divergence here is given by
\BE
\mr{div}\{\D^{ext}(\R;\theta)\}=\sum_{i=1}^{n_1} \sum_{j=1}^{n_2}\frac{\partial[\D^{ext}(\R;\theta)]_{i,j}}{\partial R_{i,j}}.
\EE
\ES
It's clear that the divergence of $\D^{ext}(\R;\theta)$ is zero. From \eqref{Eqn:SURE_DF_M}, the SURE of the MSE is given by
\BS \label{Eqn:SURE_SVT}
\begin{align}
\widehat{\mr{MSE}}&=\frac{1}{n}\|\D^{ext}(\R;\theta)-\R\|_F^2-\tau^2\\
&=\frac{1}{n}\|c\U\!\text{diag}\!\left(\bm{\Phi}(\bm{\bm{\sigma}};\theta)\right)\! \V^T\!\!-\!\U\bm{\Sigma}\V^T\|_F^2\!-\!\tau^2\\
&=\frac{1}{n}\|c\bm{\Phi}(\bm{\sigma};\theta)-\bm{\sigma}\|_2^2-\tau^2.
\end{align}
\ES
The optimal $c$ that minimizes $\widehat{\mr{MSE}}$ in (\ref{Eqn:SURE_SVT}) is given by
\BE
c^{opt}=\frac{\bm{\sigma}^T\bm{\Phi}(\bm{\sigma};\theta)}{\bm{\Phi}(\bm{\sigma};\theta)^T\bm{\Phi}(\bm{\sigma};\theta)}.
\EE
By substituting $c^{opt}$ into \eqref{Eqn:SURE_SVT}, and after some straightforward manipulations, we obtain
\BE
\widehat{\mr{MSE}}=\frac{1}{n}\left(-\frac{(\bm{\Phi}(\bm{\sigma};\theta)^T \bm{\sigma})^2}{\|\bm{\Phi}(\bm{\sigma};\theta)\|_2^2}+\|\bm{\sigma}\|_2^2\right)-\tau^2.\label{hatmse}
\EE
The optimal threshold $\theta$ that minimizes $\widehat{\mr{MSE}}$ given in (\ref{hatmse}) can be obtained by solving the following optimization problem:
\BS
\begin{align}
		\max_{\theta}&\!\quad \frac{(\bm{\Phi}(\bm{\sigma};\theta)^T \bm{\sigma})^2}{\|\bm{\Phi}(\bm{\sigma};\theta)\|_2^2}\\
		{s.t. }&\!\quad\bm{\Phi}(\bm{\sigma};\theta)\!=\!(\bm{\sigma}-\theta)_{+}\!-\!\frac{1}{n}\mr{div}\{\mr{SVT}(\R;\theta)\}\bm{\sigma}.
\end{align}\label{optsvt}\ES
The problem in (\ref{optsvt}) is non-convex. However, since only one parameter $\theta\in[0,\max(\bm{\sigma})]$ is involved, we can solve (\ref{optsvt}) by exhaustive search.
\subsection{Other Extrinsic Denoisers}
Both the SURE-LET denoiser and the SVT denoiser have analytical expressions. However, there are other denoisers that can not be expressed in a closed form. The corresponding extrinsic denoisers also have no analytical expressions. We give two examples as follows.

%
The first example is the dictionary learning denoiser. Dictionary learning aims to find a sparse representation for a given data set in the form of a linear combination of a set of basic elements. This set of basic elements is called a dictionary. Existing dictionary learning algorithms include K-SVD \cite{aharon2006img}, iterative least square (ILS) \cite{engan2007family}, recursive least squares (RLS) \cite{skretting2010recursive}, and the sequential generalization
of $K$-means (SGK) \cite{sahoo2013dictionary}. Based on above dictionary learning algorithms, we can construct dictionary learning denoisers by following the approach in \cite{elad2006image}. Specifically, consider a noisy image matrix $\R\in \mathbb{R}^{n_1\times n_2}$, where $n_1$ and $n_2$ are integers. We reshape $\R$ into a vector $\r\in \mathbb{R}^{n \times 1}$, where $n=n_1 n_2$. Also, we divide the whole image into blocks of size $n_3\times n_3$, and reshape each block $\R_{i,j}$ into a vector $\r_{i,j}$, where $n_3$ is an integer satisfying $n_3\ll n_1,n_2$. Note that $\r_{i,j}$ is related to $\r$ by $\r_{i,j}=\Ee_{i,j} \r$ where $\Ee_{i,j} \in R^{n_3^2\times n}$ is the corresponding block extraction matrix. Then we use $\{\r_{i,j}\}$ as the training set to train a dictionary $\Q\in {n_3^2\times n_4}$ using any of the dictionary learning algorithms mentioned above, where $n_4$ is an integer satisfying $n_4>n_3^2$. The image block $\r_{i,j}$ can be expressed approximately as
\BE
\r_{i,j}=\Q \Alpha_{i,j},
\EE
where $\Alpha_{i,j}\in \mathbb{R}^{n_4\times 1}$ is the sparse representation of $\r_{i,j}$ using the dictionary $\Q$. Then, we update the whole image vector $\r$ based on the learned dictionary $\Q$ and coefficients $\Alpha_{i,j}$ by averaging the denoised image block vectors as
 \begin{align}
	\begin{split}
		\tilde{\r}{=}\left(\lambda \I {+}\sum_{i,j}\Ee_{i,j}^T \Ee_{i,j} \right)^{-1}\left(\lambda \r{+}\sum_{i,j}\Ee_{i,j}^T \Q\Alpha_{i,j} \right),
	\end{split}
\end{align} 
where $\lambda$ is a constant depending on the input noise level. Finally, we reshape the image vector $\tilde{\r}$ back into an image matrix.

%

The second example is the BM3D denoiser \cite{dabov2006image}. The denoising process of BM3D is summarized as follows. First, the image matrix $\R$ is separated into image blocks of size $s_1\times s_1$ (with $7\leq s_1\leq 13$). For each image block, similar blocks are found and grouped together into a three-dimensional (3D) data array. Then, collaborative filtering is used to denoise the 3D data arrays. The filtered blocks are then returned back to their original positions. Note that BM3D achieves the state-of-the-art visual quality among all the existing image denoisers.



The above dictionary learning and BM3D denoisers have no close-form expressions, and so the divergences of these denoisers can not be calculated explicitly. Instead, we evaluate their divergences using the Monte Carlo method. Specifically, the divergence of $\D(\R)$ can be estimated by
 \begin{align} \label{Eqn:divergence_app}
 	\begin{split}
 	 \mr{div}\left\{\D(\R)\right\}\!\approx \!E_{\tilde{\bm{N}}}\!\left[\left<\!\tilde{\bm{N}},\left(\!\frac{\D(\R+\delta \tilde{\bm{N}}){-}\D(\R)}{\delta}\!\right)\!\right>\right]\!,
	\end{split}
\end{align}
where $\delta$ is a small constant, $\tilde{\bm{N}}\in \mathbb{R}^{n_1\times n_2}$ is a perturbation matrix with the elements i.i.d. drawn from $\mathcal{N}(0,1)$, and $\left<\A,\bm{B}\right>=\sum_{i,j} A_{i,j}B_{i,j}$ with $A_{i,j}$ and $B_{i,j}$ be the $(i,j)$th elements of $\A$ and $\bm{B}$, respectively. The expectation in \eqref{Eqn:divergence_app} can be approximated by sample average. It is observed in \cite{metzler2016denoising} that one sample is good enough for high-dimensional problems.
\section{Evolution Analysis of D-Turbo-CS}\label{Sec:SE-D-TurboCS}
\subsection{MSE Evolution}
The  behavior of D-Turbo-CS can be characterized by the so-called MSE evolution. Denote the input normalized mean square error (NMSE) of Module A (or equivalently, the output NMSE of Module B) at iteration $t$ as $v(t)$, and the output NMSE of Module A (or equivalently, the input NMSE of Module B) at iteration $t$ as $\tau^2(t)$, where NMSE is defined by
\begin{align}
	\begin{split}
		\text{NMSE}&=\frac{\|\hat \x-\x\|_2^2}{\|\x\|_2^2}.
	\end{split}
\end{align}
Then, the MSE evolution is characterized by 
\BS\begin{align}
\tau^2(t) &=\left(\frac{n}{m}-1\right)v(t)+\frac{n}{m}\sigma^2\\
v(t+1)&=\frac{1}{n}\E\left[ \left\| \D^{ext}\left(\bm{x}+\tau(t)\bm{e}\right)-\bm{x}\right\|_2^2\right],
\end{align}\label{Equ:SE}\ES
where the (\ref{Equ:SE}a) follows from Line 3 of Algorithm \ref{algorithm2}, (\ref{Equ:SE}b) follows from the assumption in (\ref{xbpri}), the expectation in (\ref{Equ:SE}b) is taken over $\bm{e}\sim\mathcal{N}(\bm{0},\bm{I})$, and $v(0)$ is initialized as $\E[\|\bm{x}\|_2^2]/n$. We next examine the accuracy of the above MSE evolution.

\begin{figure}[!ht]
\includegraphics[width=\columnwidth]{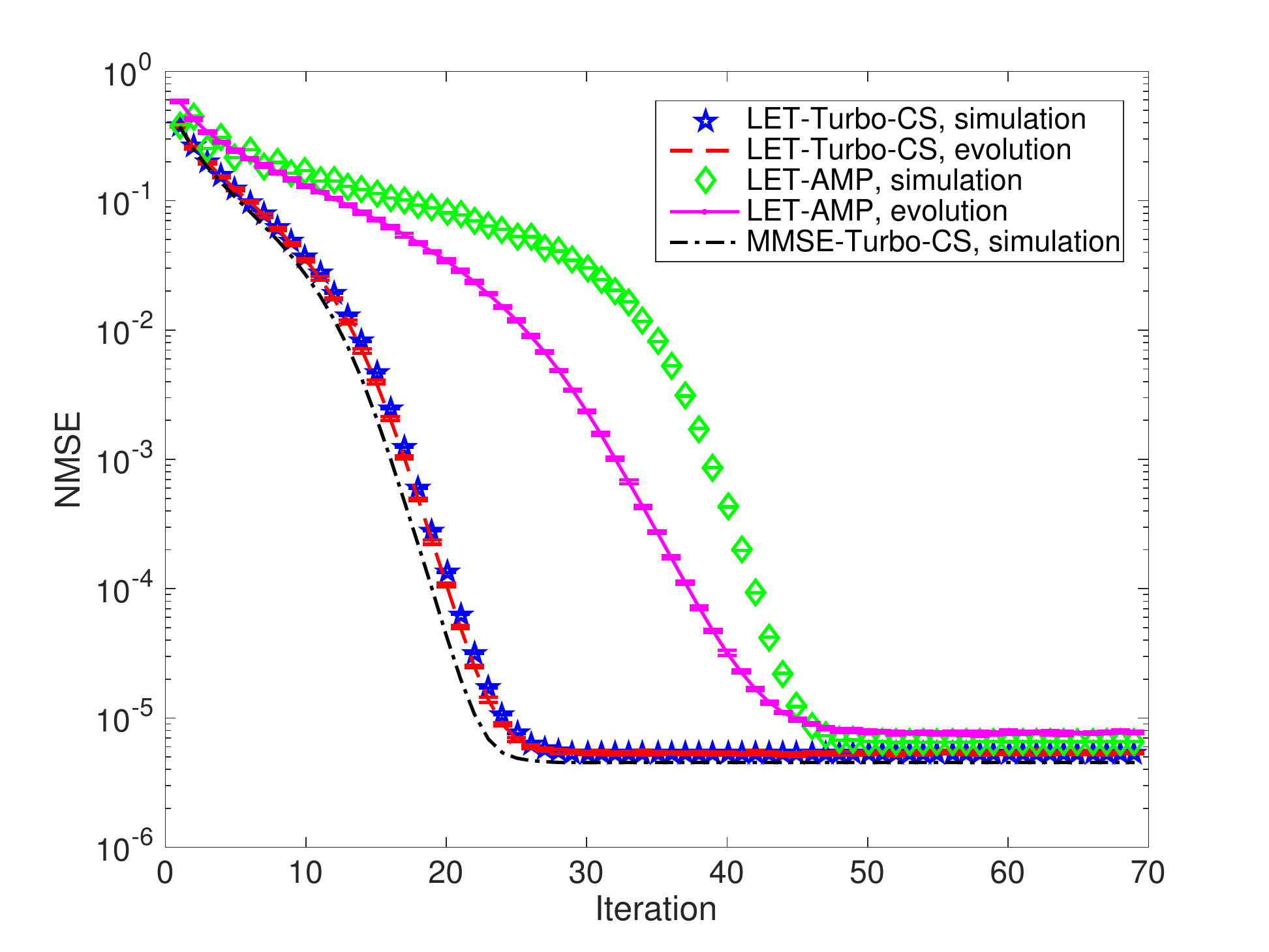}
\caption{
The MSE comparison of LET-Turbo-CS and LET-AMP with the sensing matrix given by (\ref{rpdct}).}\label{fig:figure3}
\end{figure}

\begin{figure}[!ht]
 \includegraphics[width=\columnwidth]{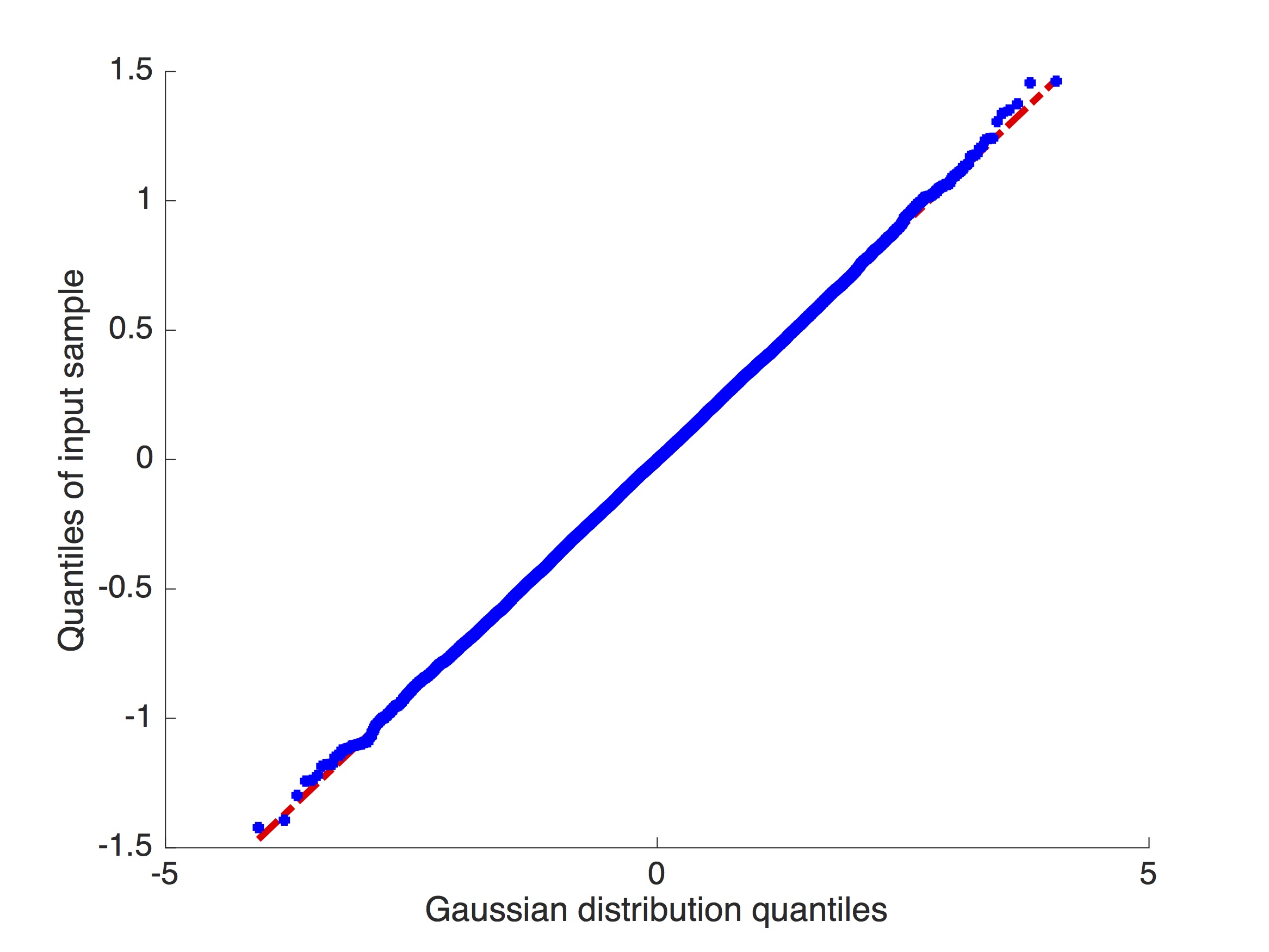}
\caption{The QQplot of the estimation error $\x_B^{pri}-\x$ in the $10$th iteration of the LET-Turbo-CS algorithm with the sensing matrix given by (\ref{rpdct}).}\label{qqplot1}
\end{figure}

\subsection{$\x$ with i.i.d. Entries}
We consider the situation of $\x$ with i.i.d. entries. In simulation, each $x_i$ in $\x$ is Gaussian-Bernoulli distributed with probability density function $p(x_i)=(1-\rho)\delta(x_i)+\rho \mathcal{N}(x_i,0,1/\rho)$, where $\delta(\cdot)$ is the Dirac delta function. The other settings are: the sparsity rate $\rho=0.27$, the measurement rate $m/n=0.5$, the signal length $n=20000$, and the sensing matrix is chosen as the random partial DCT defined by
\begin{align}
	\begin{split}
		\A_1 &=\S \W\\
	\end{split}\label{rpdct}
\end{align}
where $\S\in \mathbb{R}^{m\times n}$ is a random row selection matrix which consists of randomly selected rows from a permutation matrix, and $\W\in  \mathbb{R}^{n\times n}$ is the DCT matrix.
In simulation, the SURE-LET denoiser with the kernel functions given in (\ref{kernel1})-(\ref{kernel3}) is employed in D-Turbo-CS and D-AMP, with the corresponding algorithms denoted by LET-Turbo-CS and LET-AMP, respectively.

As shown in Fig. \ref{fig:figure3}, the MSE evolution of LET-Turbo-CS matches well with the simulation. In contrast, for LET-AMP, the state evolution deviates from the simulation. Also, LET-Turbo-CS outperforms LET-AMP\footnote{Note that, the performance of LET-AMP here is better than the original LET-AMP \cite{guo2015near}, because under the condition, LET-AMP diverges, and we replace the estimated variance $\hat\sigma^2$ in D-AMP with a more robust estimate $\hat\sigma^2=\sqrt{\frac{1}{\ln 2}}\mr{median}(|\hat x|)$ given in \cite{anitori2013design}.} considerably and performs close to MMSE-Turbo-CS in which the MMSE denoiser is employed. We also plot the QQplot of the estimation error of $\x_B^{pri}$ at iteration 10 of LET-Turbo-CS in Fig. \ref{qqplot1}. From the QQplot, we see that $\x_B^{pri}-\x$ is close to zero-mean Gaussian, which agrees well with the assumption in (\ref{xbpri}). Later, we will see that the Gaussianity of  $\x_B^{pri}-\x$ is a good indicator of the accuracy of the MSE evolution.


\begin{figure}[t!]
\includegraphics[width=\columnwidth]{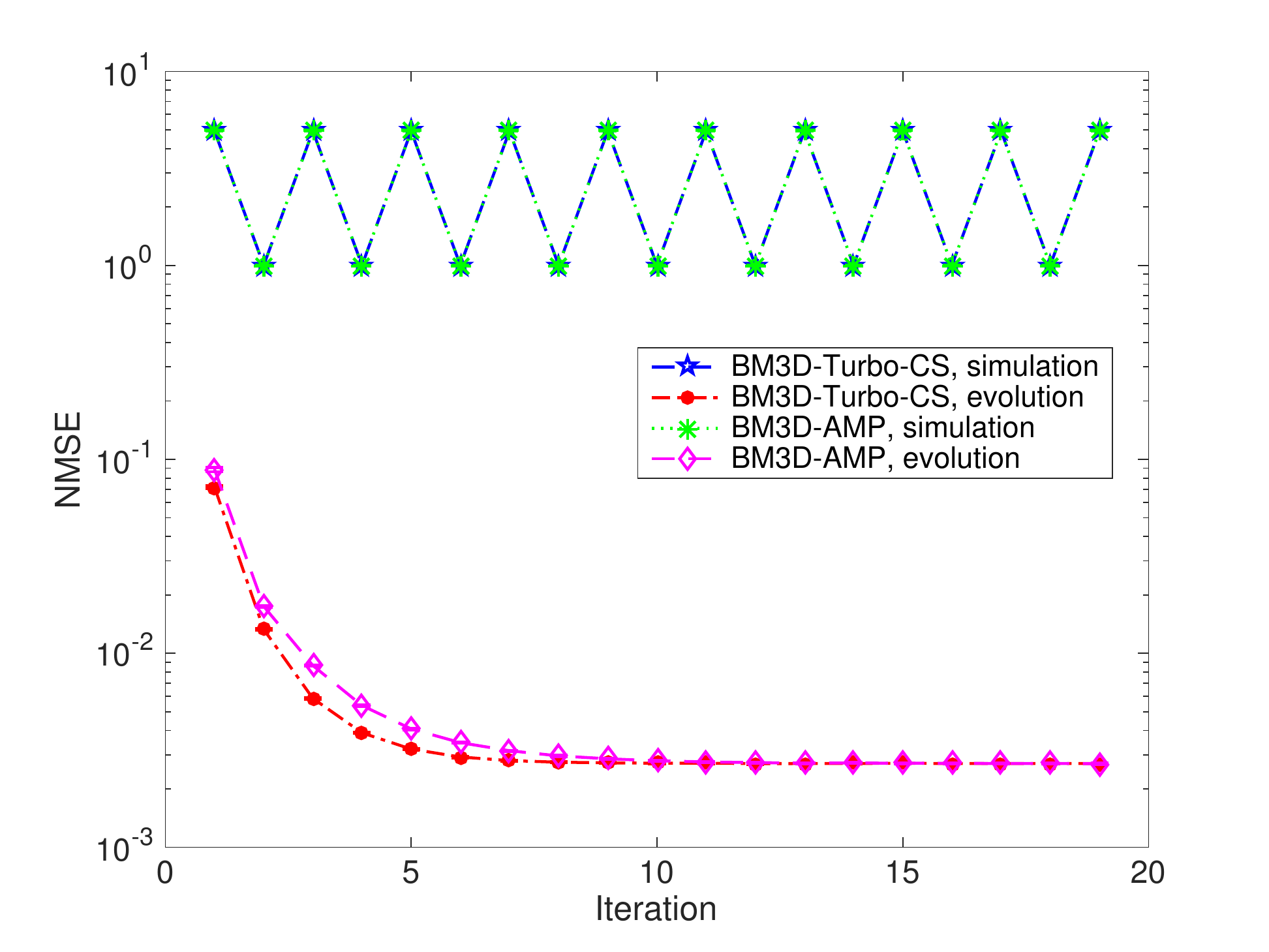}
\caption{The MSE comparison of BM3D-Turbo-CS and BM3D-AMP with the sensing matrix given by (\ref{rpdct}).}\label{fig:DDCT_matrix}
\end{figure}

\begin{figure}[t!]
 \includegraphics[width=\columnwidth]{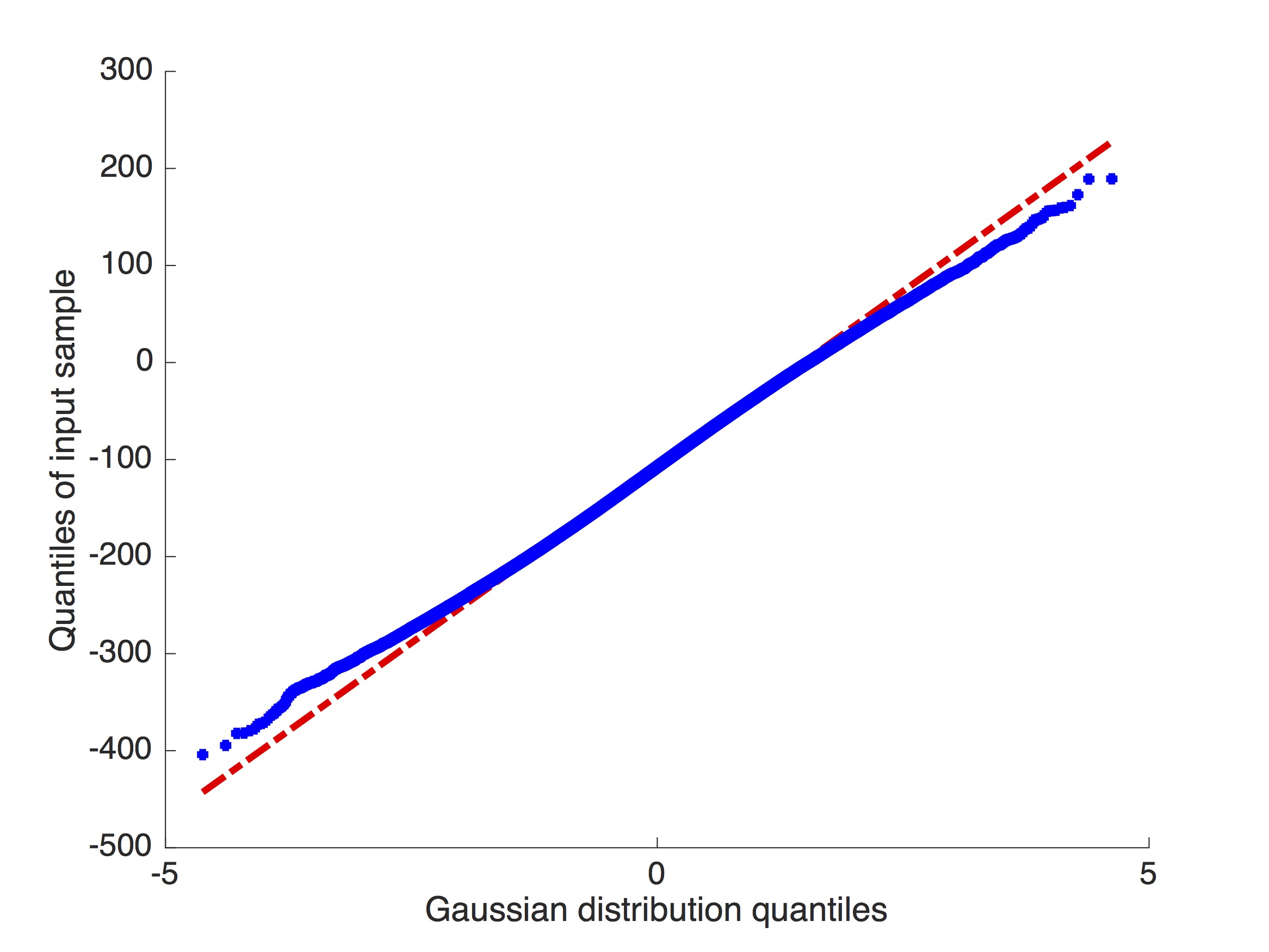}
\caption{The QQplot of the estimation error $\x_B^{pri}-\x$ in the 2nd iteration of the BM3D-Turbo-CS algorithm with the sensing matrix given by (\ref{rpdct}).}\label{qqplot3}
\end{figure}
\subsection{$\x$ with Correlated Entries}
In many applications, signals are correlated and the prior distribution is unknown. For example, the adjacent pixels of a natural image are correlated and their distributions are not available. We next study the MSE evolution of D-Turbo-CS for compressive image recovery.

In simulation, we generate signal $\x$ from the image ``Fingerprint" of size $512\times 512$ taken from the Javier Portilla’s dataset \cite{CIVtest} by reshaping the image into a vector of size $262144\times 1$. The denoiser is chosen as the BM3D denoiser, and the corresponding algorithms are denoted as BM3D-Turbo-CS and BM3D-AMP. We set the measurement rate $m/n$ to 0.3.

With the sensing matrix given in (\ref{rpdct}), the performance of BM3D-Turbo-CS and BM3D-AMP is simulated and shown in Fig. \ref{fig:DDCT_matrix}. We see that the simulation results of both algorithms do not match with the MSE evolution. Also, we plot the QQplot of the estimation error $\x_B^{pri}-\x$ in Fig. \ref{qqplot3}. We see that the distribution of $\x_A^{pri}-\x$ is not quite Gaussian, and the mean of the distribution is not zero. This interprets the failure of the evolution prediction.

\begin{figure}[!ht]
\includegraphics[width=\columnwidth]{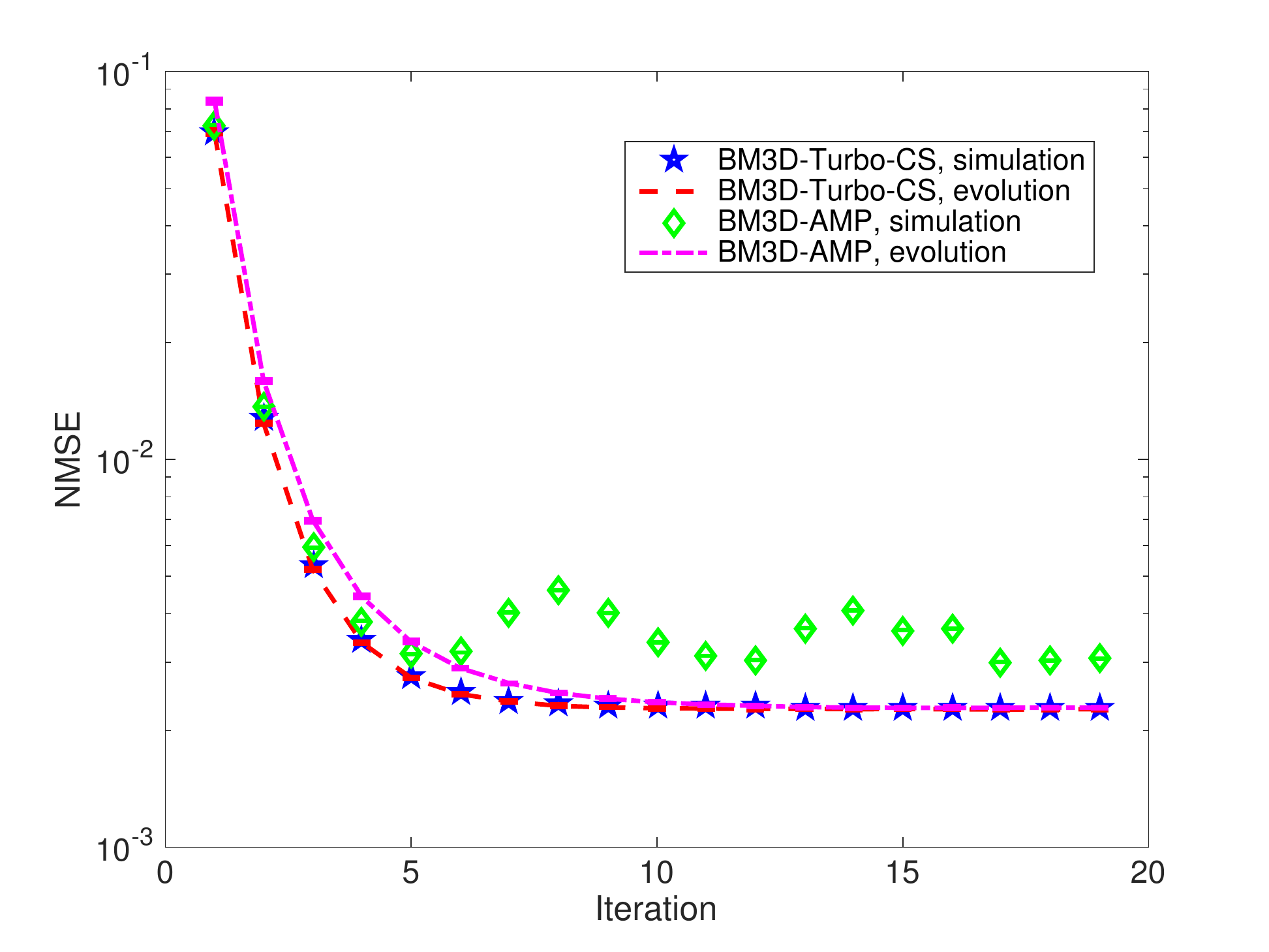}
\caption{The MSE comparison of BM3D-Turbo-CS and BM3D-AMP with the sensing matrix given by (\ref{rmdct}).}\label{fig:figure4}
\end{figure}

\begin{figure}[!ht]
 \includegraphics[width=\columnwidth]{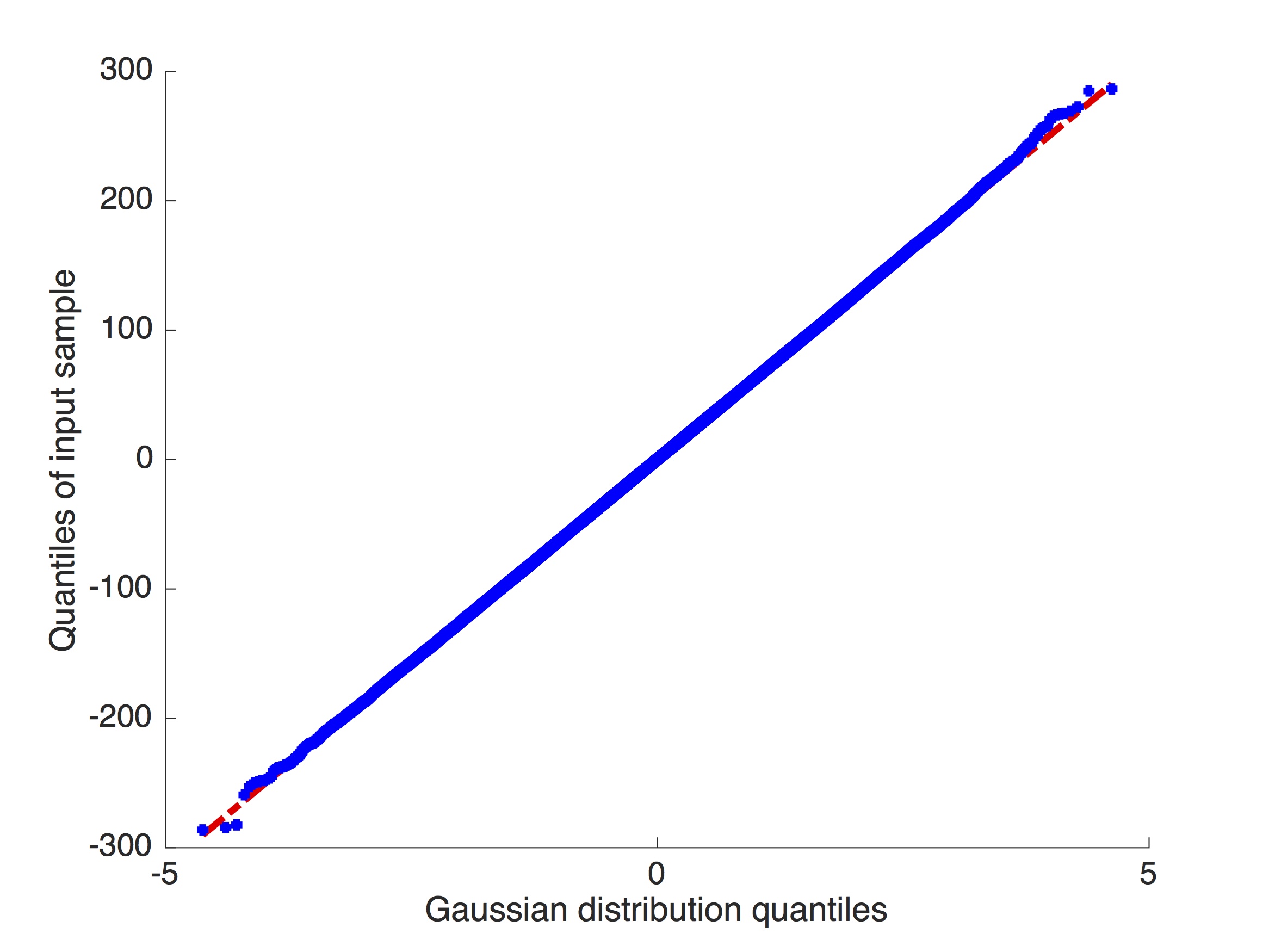}
\caption{The QQplot of the estimation error $\x_B^{pri}-\x$ in the 2th iteration of the LET-Turbo-CS algorithm with the sensing matrix given by (\ref{rmdct}).}\label{qqplot2}
\end{figure}

We conjecture that the reason for the degradation of the simulation performance in Fig. \ref{fig:DDCT_matrix} is that the correlation in $\x$ is not appropriately handled. So, we replace $\A_1$ by
\begin{align}
	\begin{split}
		\A_2 &=\S  \W\bm{\Theta}
	\end{split}\label{rmdct}
\end{align}
where $\bm{\Theta}$ is a diagonal matrix with the random signs (1 or -1) in the diagonal. The simulation result with sensing matrix $\A_2$ is shown in Fig. \ref{fig:figure4}. We see that now, the MSE evolution of BM3D-Turbo-CS matches well with the simulation. Also, BM3D-Turbo-CS outperforms BM3D-AMP in both converge rate and recovery quality. In Fig \ref{qqplot2}, the QQplot of the estimation error of $\x_B^{pri}$ at iteration 2 of BM3D-Turbo-CS is plotted. We see that the estimation error is close to zero-mean Gaussian, similarly to the case of i.i.d. $\x$. To summarize, the sensing matrix in (\ref{rpdct}) is good for i.i.d. $\x$, while the sensing matrix in (\ref{rmdct}) is needed when the entries of $\x$ are correlated.
\section{Performance Comparisons}\label{Sec:numerical}
\begin{table*}[!ht]
\centering
\begin{tabular}{l|llllll|llllll}
\hline
\textbf{Image Name} & \multicolumn{6}{c|}{\textbf{Lena}} & \multicolumn{6}{c}{\textbf{Boat}} \\ \hline
    \textbf{Measurement rate}  & \multicolumn{1}{c}{\textbf{5\%}} & \multicolumn{1}{c}{\textbf{10\%}} & \multicolumn{1}{c}{\textbf{20\%}} & \multicolumn{1}{c}{\textbf{30\%}} & \multicolumn{1}{c}{\textbf{50\%}} & \multicolumn{1}{c|}{\textbf{70\%}} &  \multicolumn{1}{c}{\textbf{5\%}}& \multicolumn{1}{c}{\textbf{10\%}} & \multicolumn{1}{c}{\textbf{20\%}} &\multicolumn{1}{c}{\textbf{30\%}} & \multicolumn{1}{c}{\textbf{50\%}} & \multicolumn{1}{c}{\textbf{70\%}} \\ \hline
EM-GM-AMP \cite{vila2013expectation} &21.56 &23.33 &25.22 & 26.89                              & 29.50                              & 32.38          &19.55 &21.06 &22.87                   & 24.70                              & 27.78                              & 30.95     \\\hline
LET-AMP \cite{guo2015near}   &-&- &- &22.09           & 31.38                              & 34.57               &- &-  &0.23          &20.02               & 29.62                              & 33.28      \\
LET-Turbo-CS   &\textbf{22.27} &\textbf{24.32} &\textbf{26.77} & \textbf{28.57}                              & \textbf{31.74}                              & \textbf{35.48}       &\textbf{19.96} &\textbf{21.86} &\textbf{24.43}                      & \textbf{26.51}                              & \textbf{30.16}                              & \textbf{34.22}  \\\hline
  BM3D-AMP \cite{metzler2016denoising}  &28.44 &31.77 &33.90 & 34.36                              & 38.55                              & 39.48         &26.15 &28.83 &31.67 & 33.54                              & 35.46                              & 39.21        \\
BM3D-Turbo-CS  &\textbf{29.28} &\textbf{31.88}&\textbf{34.35}   & \textbf{35.88}           & \textbf{38.60}                              & \textbf{42.64}          &\textbf{26.15} &\textbf{29.03}&\textbf{32.32}   & \textbf{34.30}           & \textbf{37.32}                              & \textbf{40.92}             \\ \hline
  SGK-AMP \cite{Li2016}  &7.67 &8.27 &27.92  & 29.85                                  & 33.17                              & 35.87           &5.35&5.53 &25.49                  & 28.07                                  & 31.42                              & 34.57       \\
SGK-Turbo-CS   &\textbf{7.70} &\textbf{8.35} &\textbf{29.01}  & \textbf{31.30}                              & \textbf{34.60}                              & \textbf{37.85}              &\textbf{5.39} &\textbf{5.56} &\textbf{26.22}               & \textbf{28.85}                              & \textbf{32.54}                              & \textbf{35.90}            \\ \hline
\textbf{Image Name} & \multicolumn{6}{c|}{\textbf{Barbara}} & \multicolumn{6}{c}{\textbf{Fingerprint}}                                                                   \\ \hline
 \textbf{Measurement rate}  & \multicolumn{1}{c}{\textbf{5\%}}& \multicolumn{1}{c}{\textbf{10\%}} & \multicolumn{1}{c}{\textbf{20\%}} & \multicolumn{1}{c}{\textbf{30\%}} & \multicolumn{1}{c}{\textbf{50\%}} & \multicolumn{1}{c|}{\textbf{70\%}} & \multicolumn{1}{c}{\textbf{5\%}}&\multicolumn{1}{c}{\textbf{10\%}} & \multicolumn{1}{c}{\textbf{20\%}}&\multicolumn{1}{c}{\textbf{30\%}} & \multicolumn{1}{c}{\textbf{50\%}} & \multicolumn{1}{c}{\textbf{70\%}} \\ \hline
EM-GM-AMP \cite{vila2013expectation}    &18.54 &20.56 &22.65  &24.47                              & 27.69                              & 32.14        &16.81 &18.24 &20.37                      & 22.51                              & 26.04                              &29.58                         \\\hline
LET-AMP \cite{guo2015near} &- &- &- &19.92                        & 27.57                              & 31.15     &-  &- &-                        &17.78                   & 29.05                              & 33.46                        \\
LET-Turbo-CS &\textbf{18.87} &\textbf{20.65} &\textbf{22.86}   & \textbf{24.58}                              & \textbf{28.07}                              & \textbf{32.36}        &\textbf{16.03} &\textbf{18.03} &\textbf{22.09}                       & \textbf{24.05}                              & \textbf{29.83}                              & \textbf{34.90} \\\hline
  BM3D-AMP \cite{metzler2016denoising}  &\textbf{26.74} &29.52 &32.81 & 35.21                               & 38.46                             & 41.66               &18.18 &22.75 &26.61                & 28.59                                 & 32.07                              & 36.44                          \\
BM3D-Turbo-CS   &26.73 &\textbf{30.40}&\textbf{34.23}   & \textbf{36.46}                              & \textbf{39.91}                       & \textbf{43.37}        &\textbf{18.04} &\textbf{24.53} &\textbf{27.73}                      & \textbf{30.28}                              & \textbf{34.51}                              & \textbf{38.91}                 \\ \hline
  SGK-AMP \cite{Li2016} &\textbf{5.94} &6.35 &25.58 & 28.20                                 & 32.13                              & 35.78          &\textbf{4.59} &4.76&20.32                    & 23.98                                 & \textbf{28.49}                            & 32.39                          \\
SGK-Turbo-CS    &5.88 &\textbf{6.36} &\textbf{26.43}  & \textbf{29.30}                              & \textbf{33.65}                              & \textbf{37.77}           &4.58 &\textbf{4.82} &\textbf{20.46}                   & \textbf{24.10}                              & 28.38                              & \textbf{32.79}                 \\
\hline
\end{tabular}
\caption{The PSNR of the reconstructed images with sensing matrix $\A_2$.}
\label{table1}
\end{table*}

\begin{table*}[!ht]
\centering
\begin{tabular}{l|llllll|llllll}
\hline
\textbf{Image Name} & \multicolumn{6}{c|}{\textbf{Lena}}& \multicolumn{6}{c}{\textbf{Boat}}\\ \hline
\textbf{Measurement rate} & \multicolumn{1}{c}{\textbf{5\%}}& \multicolumn{1}{c}{\textbf{10\%}} & \multicolumn{1}{c}{\textbf{20\%}}    & \multicolumn{1}{c}{\textbf{30\%}} & \multicolumn{1}{c}{\textbf{50\%}} & \multicolumn{1}{c|}{\textbf{70\%}} & \multicolumn{1}{c}{\textbf{5\%}}& \multicolumn{1}{c}{\textbf{10\%}} & \multicolumn{1}{c}{\textbf{20\%}} &\multicolumn{1}{c}{\textbf{30\%}} & \multicolumn{1}{c}{\textbf{50\%}} & \multicolumn{1}{c}{\textbf{70\%}} \\\hline
LET-AMP \cite{guo2015near}   &3.14&2.43&2.57&2.41           & 2.46                            & 2.58                             &2.91&2.35&2.42&2.47              & 3.65                            &  3.37        \\
LET-Turbo-CS  &\textbf{1.40}&\textbf{1.23}&\textbf{1.23}  & \textbf{1.19}                              & \textbf{0.93}                              & \textbf{0.85}     &\textbf{0.97}&\textbf{1.16}&\textbf{1.21}                         & \textbf{1.39}                              & \textbf{1.63}                              & \textbf{1.27} \\
 \hline
\textbf{Image Name} & \multicolumn{6}{c|}{\textbf{Barbara}} & \multicolumn{6}{c}{\textbf{Fingerprint}}\\ \hline
\textbf{Measurement rate}  & \multicolumn{1}{c}{\textbf{5\%}}& \multicolumn{1}{c}{\textbf{10\%}} & \multicolumn{1}{c}{\textbf{20\%}}      & \multicolumn{1}{c}{\textbf{30\%}} & \multicolumn{1}{c}{\textbf{50\%}} & \multicolumn{1}{c|}{\textbf{70\%}} &  \multicolumn{1}{c}{\textbf{5\%}}& \multicolumn{1}{c}{\textbf{10\%}} & \multicolumn{1}{c}{\textbf{20\%}} &   \multicolumn{1}{c}{\textbf{30\%}} & \multicolumn{1}{c}{\textbf{50\%}} & \multicolumn{1}{c}{\textbf{70\%}}  \\\hline
LET-AMP \cite{guo2015near}    &3.48&3.55&3.11      &3.86                         & 3.58                              & 3.24            &4.41&3.61&3.41                 & 3.30                   & 3.10                             & 4.08                  \\
LET-Turbo-CS   &\textbf{1.44}&\textbf{1.58}&\textbf{1.37} & \textbf{1.41}                              & \textbf{1.75}                              & \textbf{1.46}              &\textbf{1.30}&\textbf{2.40}&\textbf{3.08}               & \textbf{3.02}                              & \textbf{1.69}                              & \textbf{1.46} \\
\hline
\end{tabular}
\caption{The recovery time of different images for sensing matrix $\A_2$. The unit of time is second.}
\label{table2}
\end{table*}

\begin{table*}[!ht]
\centering
\begin{tabular}{l|lll|lll|lll|lll}
\hline
\textbf{Sensing matrices} & \multicolumn{3}{c|}{\textbf{$\A_1$}}    & \multicolumn{3}{c|}{\textbf{$\A_2$}} &  \multicolumn{3}{c|}{\textbf{$\A_1$}}    & \multicolumn{3}{c}{\textbf{$\A_2$}}\\\hline
\textbf{Image Name} & \multicolumn{6}{c|}{\textbf{Lena}}& \multicolumn{6}{c}{\textbf{Boat}}\\ \hline
\textbf{Measurement rate}  & \multicolumn{1}{c}{\textbf{30\%}}& \multicolumn{1}{c}{\textbf{50\%}} & \multicolumn{1}{c|}{\textbf{70\%}}      & \multicolumn{1}{c}{\textbf{30\%}} & \multicolumn{1}{c}{\textbf{50\%}} & \multicolumn{1}{c|}{\textbf{70\%}} &  \multicolumn{1}{c}{\textbf{30\%}}& \multicolumn{1}{c}{\textbf{50\%}} & \multicolumn{1}{c|}{\textbf{70\%}} &   \multicolumn{1}{c}{\textbf{30\%}} & \multicolumn{1}{c}{\textbf{50\%}} & \multicolumn{1}{c}{\textbf{70\%}}  \\\hline
BM3D-AMP \cite{metzler2016denoising}  &\textbf{7.68} &7.59 &13.83 & 34.36                              & 38.55                              & 39.48         &- &9.23 &19.11 & 33.54                              & 35.46                              & 39.21        \\
BM3D-Turbo-CS  &\textbf{7.68}&\textbf{8.69}&\textbf{13.84}   & \textbf{35.88}           & \textbf{38.60}                              & \textbf{42.64}          &- &\textbf{9.32}&\textbf{19.15}   & \textbf{34.30}           & \textbf{37.32}                              & \textbf{40.92}             \\ \hline
\textbf{Image Name} & \multicolumn{6}{c|}{\textbf{Barbara}} & \multicolumn{6}{c}{\textbf{Fingerprint}}\\ \hline
\textbf{Measurement rate}  & \multicolumn{1}{c}{\textbf{30\%}}& \multicolumn{1}{c}{\textbf{50\%}} & \multicolumn{1}{c|}{\textbf{70\%}}      & \multicolumn{1}{c}{\textbf{30\%}} & \multicolumn{1}{c}{\textbf{50\%}} & \multicolumn{1}{c|}{\textbf{70\%}} &  \multicolumn{1}{c}{\textbf{30\%}}& \multicolumn{1}{c}{\textbf{50\%}} & \multicolumn{1}{c|}{\textbf{70\%}} &   \multicolumn{1}{c}{\textbf{30\%}} & \multicolumn{1}{c}{\textbf{50\%}} & \multicolumn{1}{c}{\textbf{70\%}}  \\\hline
 BM3D-AMP \cite{metzler2016denoising}  &\textbf{5.88} &9.93 &21.06 & 35.21                               & 38.46                             & 41.66               &- &7.9 &20.05                &      28.59  & 32.07       & 36.44                    \\
BM3D-Turbo-CS   &\textbf{5.88} &\textbf{10.18}&\textbf{21.07}   & \textbf{36.46}                              & \textbf{39.91}                       & \textbf{43.37}        &-&\textbf{8.39}&\textbf{20.07}                     &\textbf{30.28}                              & \textbf{34.51}                              & \textbf{38.91}             \\ \hline
\end{tabular}
\caption{The PSNR of the reconstructed images for different sensing matrices.}
\label{table3}
\end{table*}

In this section, we provide numerical results of D-Turbo-CS for compressive image recovery and low-rank matrix recovery. For comparison, the recovery accuracy is measured by peak signal-to-noise ratio (PSNR):
\begin{align}
	\text{PNSR}&=10\log_{10}\left(\frac{\text{MAX}^2}{\text{MSE}}\right),
\end{align}
where MAX denotes the maximum possible pixel value of the image.

The stopping criterion of D-Turbo-CS is described as follows. The D-Turbo-CS algorithm stops when its output at iteration $t$ $\hat\x(t)$ satisfies $\frac{\|\hat \x(t)-\hat \x(t-1)\|^2}{\|\hat \x(t-1)\|^2}\leq \epsilon$ or when it is excecuted for over $T$ iterations, where $\epsilon$ and $T$ are predetermined constants.

\begin{figure}[!ht]
\centering
\includegraphics[width=0.75\columnwidth]{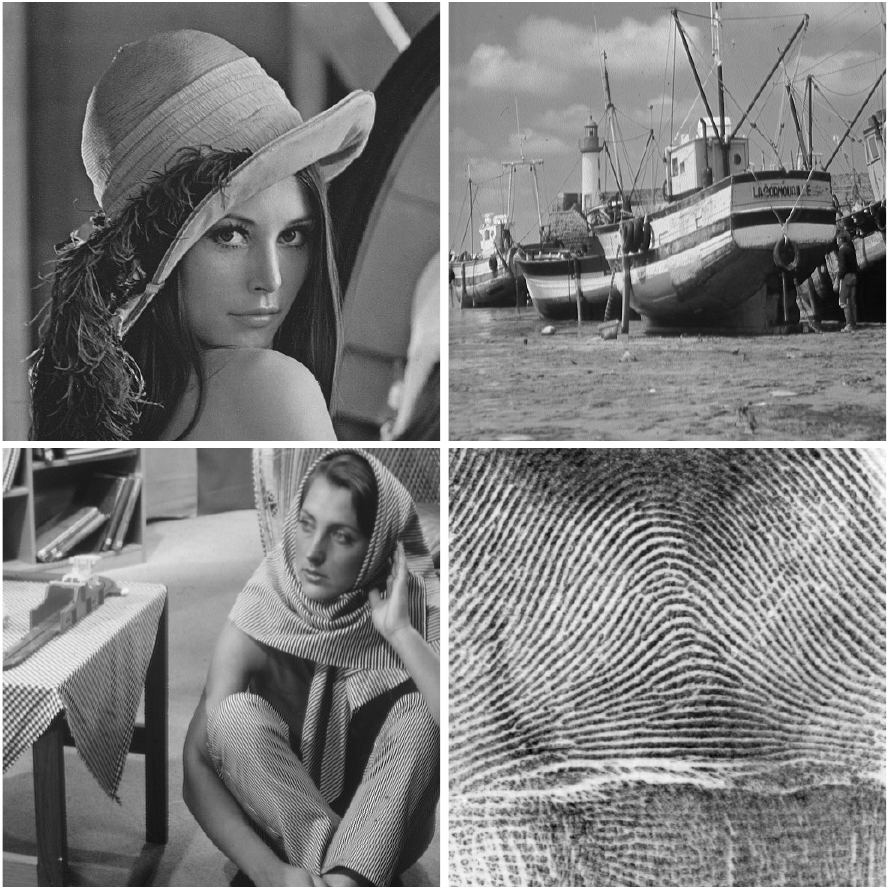}
\caption{The test images of size $512\times 512$.}\label{fig:imgs}
\vspace{-1em}
\end{figure}


\subsection{Noiseless Image Recovery}
For noiseless compressive image recovery, we consider three denoisers mentioned in Section \ref{denoisers}: the SURE-LET denoiser, the BM3D denoiser and the dictionary learning denoiser. The corresponding algorithms of D-Turbo-CS and D-AMP are denoted by LET-Turbo-CS and LET-AMP, BM3D-Turbo-CS and BM3D-AMP, and SGK-Turbo-CS and SGK-AMP. The EM-GM-AMP algorithm in \cite{vila2013expectation} is also included for comparison. The test images are chosen from the Javier Portilla’s dataset, including ``Lena", ``Boat", ``Barbara" and, ``Fingerprint" in Fig. \ref{fig:imgs}. The settings of $\epsilon$ and $T$ are as follows: $\epsilon=10^{-4}$ and $T=20$ for SURE-LET; $\epsilon=10^{-4}$ and $T=30$ for BM3D; $\epsilon=10^{-4}$ and $T=20$ for SGK.

In Table \ref{table1}, we compare D-Turbo-CS with D-AMP and EM-GM-AMP for noiseless natural image recovery with the sensing matrix given in (\ref{rmdct}). We see that D-Turbo-CS outperforms D-AMP and EM-GM-AMP for all the test images under almost all measurement rates and denoisers. To compare the reconstruction speed, we further report the reconstruction time of LET-AMP and LET-Turbo-CS in Table \ref{table2}. Both algorithms are run until the stopping criterion is activated. We see that the reconstruction time of LET-Turbo-CS is much less than that of LET-AMP. In Table \ref{table3}, we list the PSNR of reconstructed images using BM3D-AMP and BM3D-Turbo-CS for sensing matrix $\A_1$ and $\A_2$. From the table, we see that the recovery quality for sensing matrix $\A_1$ is very poor, which is consistent with the observation in Fig. \ref{fig:DDCT_matrix}. To summarize, D-Turbo-CS has significant advantages over D-AMP and EM-GM-AMP in compressive image recovery in both visual quality and recovery time.

 \begin{figure}[!ht]
 \includegraphics[width=\columnwidth]{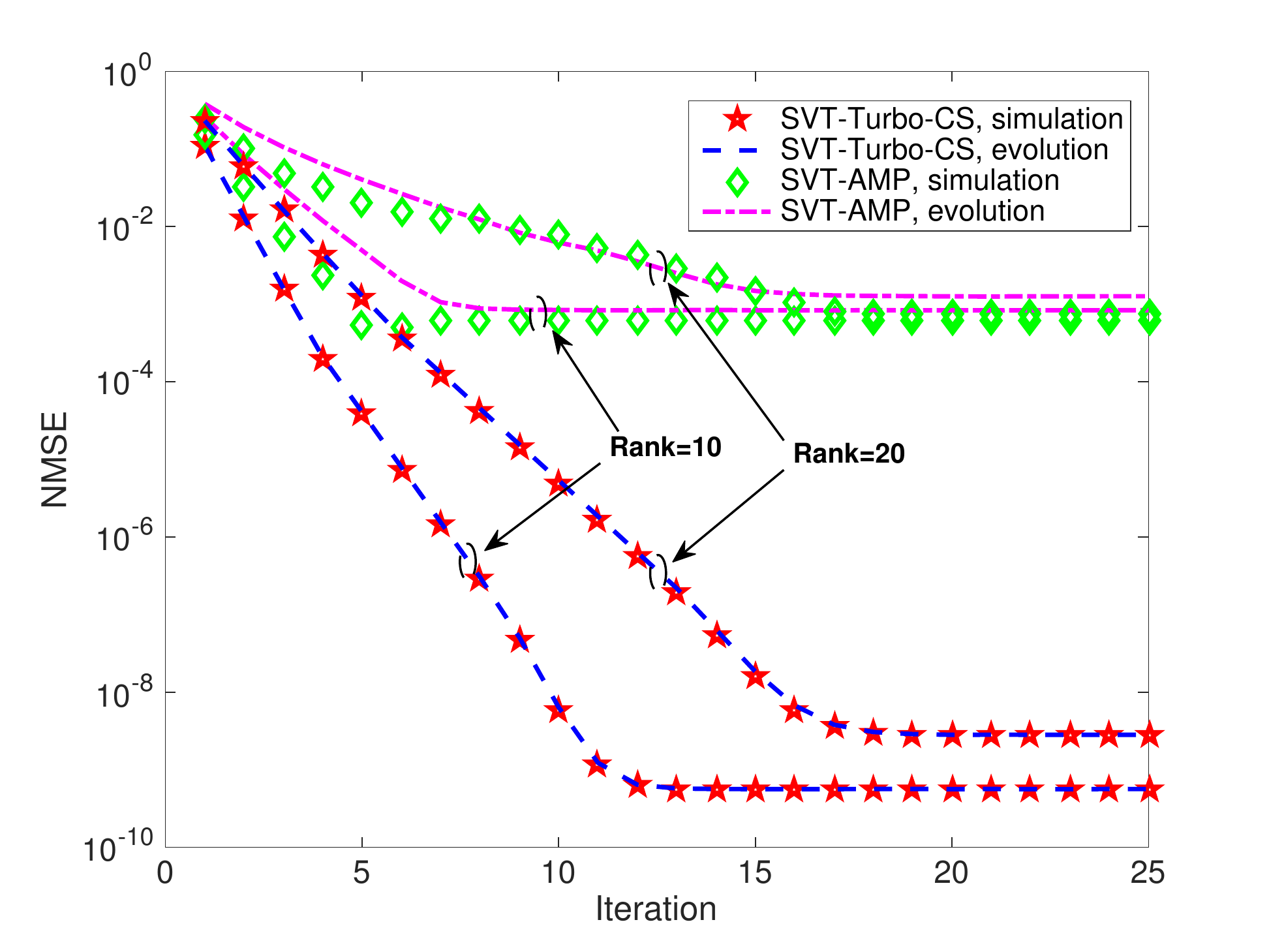}
 \caption{
 The performance comparison of SVT-Turbo-CS and SVT-AMP for low-rank matrix recovery with the sensing matrix $\A_2$.}\label{fig:figure8}
 \end{figure}
 
\subsection{Low-Rank Matrix Recovery}
For low-rank matrix recovery, we use the SVT denoiser. The corresponding algorithms of D-Turbo-CS and D-AMP are denoted respectively by SVT-Turbo-CS and SVT-AMP. The low-rank matrix $\X$ is generated by the multiplication of two random matrices of size $128\times 10$ and $10\times 128$, with the elements of the two matrices independently drawn from $\mathcal{N}(0,1)$.

The NMSE comparison of SVT-Turbo-CS and SVT-AMP under the measurement rate $\delta=m/n=0.48$ with sensing matrix given in (\ref{rmdct}) is shown in Fig. \ref{fig:figure8}. We see that, SVT-Turbo-CS significantly outperforms SVT-AMP, and the MSE evolution of SVT-Turbo-CS agrees well with the simulation result.


\section{Conclutions}\label{conclusion}
In this paper, we developed the D-Turbo-CS algorithm for compressed sensing. We discussed how to construct and optimize the so-called extrinsic denoisers for D-Turbo-CS. D-Turbo-CS does not require prior knowledge of the signal distribution, and so can be adopted in many applications including compressive image recovery and low-rank matrix recovery. Numerical results show that D-Turbo-CS outperforms D-AMP and EM-GM-AMP in terms of both recovery accuracy and convergence speed when partial orthogonal sensing matrices are involved.



%

%
%
%
%
%

\ifCLASSOPTIONcaptionsoff
  \newpage
\fi

\bibliographystyle{IEEEtran}	

\end{document}